\documentclass[twocolumn]{aastex63}
\submitjournal{ApJ}
\shorttitle{ALMA polarization observations in an asymmetric dust trap}
\shortauthors{Ohashi et al.}
\graphicspath{{./}{figures/}}
\begin{document}

\title{Solving grain size inconsistency between ALMA polarization and VLA continuum in the Ophiuchus IRS 48 protoplanetary disk}

\author[0000-0002-9661-7958]{Satoshi Ohashi}
\affil{RIKEN Cluster for Pioneering Research, 2-1, Hirosawa, Wako-shi, Saitama 351-0198, Japan}
\email{satoshi.ohashi@riken.jp}

\author[0000-0003-4562-4119]{Akimasa Kataoka}
\affiliation{National Astronomical Observatory of Japan, 2-21-1 Osawa, Mitaka, Tokyo 181-8588, Japan}
\affiliation{Department of Astronomical Science, SOKENDAI (The Graduate University for Advanced Studies), 2-21-1 Osawa, Mitaka, Tokyo 181-8588, Japan}

\author[0000-0003-2458-9756]{Nienke Van der Marel}
\affiliation{Department of Physics and Astronomy, University of Victoria 3800 Finnerty Road, Victoria, BC, V8P 5C2, Canada}

\author[0000-0002-8975-7573]{Charles L. H. Hull}
\altaffiliation{NAOJ fellow} 
\affiliation{National Astronomical Observatory of Japan, NAOJ Chile, Alonso de C\'ordova 3788, Office 61B, 7630422, Vitacura, Santiago, Chile}
\affiliation{Joint ALMA Observatory, Alonso de C\'ordova 3107, Vitacura, Santiago, Chile}

\author{William R. F. Dent}
\affiliation{Joint ALMA Observatory, Alonso de C\'ordova 3107, Vitacura, Santiago, Chile}

\author{Adriana Pohl}
\affiliation{Max Planck Institute for Astronomy, K{\"o}nigstuhl 17, 69117 Heidelberg, Germany}

\author[0000-0001-8764-1780]{Paola Pinilla}
\affiliation{Max Planck Institute for Astronomy, K{\"o}nigstuhl 17, 69117 Heidelberg, Germany}

\author{Ewine F. van Dishoeck}
\affiliation{Leiden Observatory, Leiden University, P.O. Box 9513, NL-2300 RA Leiden, The Netherlands}

\author{Thomas Henning}
\affiliation{Max Planck Institute for Astronomy, K{\"o}nigstuhl 17, 69117 Heidelberg, Germany}

%\nocollaboration{2}

%% Note that the \and command from previous versions of AASTeX is now
%% depreciated in this version as it is no longer necessary. AASTeX 
%% automatically takes care of all commas and "and"s between authors names.

%% AASTeX 6.3 has the new \collaboration and \nocollaboration commands to
%% provide the collaboration status of a group of authors. These commands 
%% can be used either before or after the list of corresponding authors. The
%% argument for \collaboration is the collaboration identifier. Authors are
%% encouraged to surround collaboration identifiers with ()s. The 
%% \nocollaboration command takes no argument and exists to indicate that
%% the nearby authors are not part of surrounding collaborations.

%% Mark off the abstract in the ``abstract'' environment. 
\begin{abstract}

The protoplanetary disk around Ophiuchus IRS 48 shows an azimuthally asymmetric dust distribution in (sub-)millimeter  observations, which is interpreted as a vortex, where millimeter/centimeter-sized particles are trapped at the location of the continuum peak.
In this paper, we present 860 $\mu$m ALMA observations of polarized dust emission of this disk.
The polarized emission was detected toward a part of the disk.
The polarization vectors are parallel to the disk minor axis, and the polarization fraction was derived to be $1-2$\%. These characteristics are consistent with models of self-scattering of submillimeter-wave emission, which indicate a maximum grain size of $\sim100$ $\mu$m. However, this is inconsistent with the previous interpretation of millimeter/centimeter dust particles being trapped by a vortex.
To explain both, ALMA polarization and previous ALMA and VLA observations, we suggest that the thermal emission at 860 $\mu$m wavelength is optically thick ($\tau_{\rm abs}\sim7.3$) at the dust trap with the maximum observable grain size of $\sim100$ $\mu$m rather than an optically thin case with $\sim$ cm dust grains. 
We note that we cannot rule out that larger dust grains are accumulated near the midplane if the 860 $\mu$m thermal emission is optically thick.

\end{abstract}

%% Keywords should appear after the \end{abstract} command. 
%% See the online documentation for the full list of available subject
%% keywords and the rules for their use.

%% From the front matter, we move on to the body of the paper.
%% Sections are demarcated by \section and \subsection, respectively.
%% Observe the use of the LaTeX \label
%% command after the \subsection to give a symbolic KEY to the
%% subsection for cross-referencing in a \ref command.
%% You can use LaTeX's \ref and \label commands to keep track of
%% cross-references to sections, equations, tables, and figures.
%% That way, if you change the order of any elements, LaTeX will
%% automatically renumber them.
%%
%% We recommend that authors also use the natbib \citep
%% and \citet commands to identify citations.  The citations are
%% tied to the reference list via symbolic KEYs. The KEY corresponds
%% to the KEY in the \bibitem in the reference list below. 

\section{Introduction} \label{sec:intro}

The evolution of dust grains in protoplanetary disks is one of the most important processes for planet formation.
The dust grains are thought to grow from micron-sized particles to pebbles (millimeter/centimeter-sized particles), and beyond to planetesimals in protoplanetary disks \citep[e.g.,][]{tes14}.

Recent millimeter and submillimeter observations with the Atacama Large Millimeter/submillimeter Array (ALMA) have revealed that protoplanetary disks have a variety of structures such as rings, spirals, and lopsided structures \citep[e.g.,][]{van13,cas13,per14,alma15,and16,can16,ise16,tsu16,cie17,loo17,kra17,van17,and18,ans18,ber18,boe18,cla18,dip18,don18,fed18,lon18,she18,van18,van19}.
These structures of disks may be related to grain growth and planet formation.
In order to understand where and how dust grains grow, it is essential to measure the grain sizes in these structures.

One way to measure the size of the dust grains is to derive the frequency dependence of thermal dust continuum emission since larger grains  efficiently emit thermal radiation at a wavelength similar to their size \citep[e.g.,][]{dra06}.
Therefore, the spectral index, $\alpha$, provides us with information on grain sizes.
Multi-wavelength observations using ALMA, Submillimeter Array (SMA), Plateau de Bure Interferometer (PdBI), Nobeyama Millimeter Array (NMA) and Karl G. Jansky Very Large Array (VLA) have been used to derive spectral indices for disks in various star-forming regions.
The spectral index is derived to be $\alpha\sim2-2.5$, suggesting grain growth to sizes of $\sim1$ mm \citep[e.g.,][]{kit02,and05,ric10,ric12,tes14,ans18}.

Another recently proposed way to measure dust grain sizes is via polarization of millimeter/submillimeter-wave thermal dust continuum emission due to scattering \citep{kat15,poh16,yang16}.
As dust grains grow, the scattering opacity becomes as high as the absorption opacity at submillimeter wavelengths.  Thus, the continuum emission is expected to be polarized due to self-scattering.
The polarization fraction is the highest if dust grains have a maximum size of $a_{\rm max}\sim \lambda/2\pi$ where $\lambda$ is the observing wavelength. 

Thanks to the high spatial resolution and high sensitivity of ALMA, polarization of dust continuum emission has been detected not only in star forming regions but also in protoplanetary disks.
Several studies have shown that the polarization vectors are parallel to the disk minor axis \citep[e.g.,][]{ste14,kat17,ste17,lee18,cox18,gir18,sad18,hul18,har18,bac18,den19,har19,mor19,sad19}, which is consistent with the self-scattering model for an inclined disk \citep{yang16,yang17,kat16}.
According to the self-scattering process, these ALMA polarization observations suggest that a maximum grain size is about 100 $\mu$m, which is smaller than the sizes inferred from the spectral index.
\citet{oha19} investigated the polarization of the HD 163296 disk, which has multiple ring and gap structures, by comparing observations with radiative transfer models. They found that polarization is mainly produced in the gaps rather than in the rings, indicating different grain sizes between them. The gaps have a maximum grain size of $\sim100$ $\mu$m, while the rings have significantly larger or smaller dust grains.
In the case of larger dust grains, these rings may be due to dust traps \citep{dul18}.
Therefore, the grain size measurements from polarization observations and from spectral index studies may trace different parts of disks.

Another possibility is that lower spectral indices may be due to optically thick emission. 
Optically thick emission follows black body radiation, which means the spectral index $\alpha=2$.
By taking into account the effects of scattering, an observed spectral index can be even {\textit lower} than 2 when (sub)millimeter emission is optically thick and grain sizes are on the order of 100 micron \citep{liu19,zhu19}.
A detailed analysis of ALMA and VLA observations with absorption and scattering effects has been performed for two disks, HL Tau and TW Hya \citep{car19,ued20}.
\citet{car19} showed that dust grains have grown to a few millimeters within 20 au of the center of the HL Tau disk. However, this is inconsistent with a grain size measured to be $\sim100$ $\mu$m by the ALMA polarization observations \citep{kat17,ste17}.
\citet{lin19} also investigated the HD 163296 disk by ALMA polarization and spectral index.
They showed that the low spectral index in the rings can be explained even by the dust grains with $\sim100$ $\mu$m.
Due to the high optical depth of the rings in the sub-mm region, it is difficult to constrain the grain size.
To investigate how and when dust grains grow, it is essential to understand the different grain size measurements from the spectral index and from polarization.

The target for this study is a transition disk around Ophiuchus IRS 48. 
The Ophiuchus IRS 48 is a young A-type star with a distance of 120 pc\footnote{The distance is revised to be 134 pc by \citet{gai18}. However, we still use a a distance of 120 pc to compare with the previous observations} \citep{loi08}.
The ring structure was firstly imaged with the VISIR instrument on the Very Large Telescope (VLT) at 18.7 $\mu$m \citep{gee07}.
At longer wavelengths, \citet{van13} showed a highly asymmetric, peanut-shaped, structure by observing 440 $\mu$m dust continuum emission with ALMA.
The peak emission of this source is $\gtrsim 100$ times brighter than the opposite side of the ring.
Subsequently, \citet{van15} observed the disk with Very Large Array (VLA) 8.8 mm continuum emission and found that longer wavelength continuum emission is more concentrated on the dust trap; they assumed optically thin emission at both ALMA and VLA wavelengths. These results suggest an increase of large particles in the center of the trap.
On the other hand, the gas distribution traced by CO isotopes and the micrometer-sized dust distributions traced by mid-infrared and near-infrared scattered light emission are  axisymmeteric \citep{van13,bru14}.
The different distributions between large dust grains and small dust grains/gas have been interpreted as a dust trap caused by a vortex.
Larger dust grains are more concentrated in the dust trap \citep{bir13}.
Such dust traps caused by a vortex may be triggered by instabilities such as the Rossby wave instability \cite[e.g.,][]{lov99,lin12,lyr13,zhu14,flo15,ono16}.
Thus, the IRS 48 disk is a good test case to investigate the dust distributions from both ALMA polarization and spectral index among ALMA 440 $\mu$m, 860 $\mu$m, and VLA 8.8 mm. 
We discuss the reasonable dust model to explain the two different ways of measuring the grain size.

The rest of this paper is organized as follows. In Section \ref{sec:obs}, we describe  the ALMA polarization observation setup and calibration process. Section \ref{sec:res} presents the ALMA polarization images of the disk. We also show the VLA image taken by \citet{van15} in this section because we discuss the dust grain sizes by using both ALMA polarization and VLA continuum data.
In Section \ref{sec:model}, we perform the radiative transfer calculations by taking into account the self-scattering to compare the ALMA and VLA images and investigate the grain size.
Section \ref{sec:discussion} discusses the interpretation of both ALMA polarization and VLA continuum emission and possible implications for the dust grain sizes and its distributions.
Finally, the results and the conclusions of this work are summarized in Section \ref{sec:conclusion}.

\section{Observations}\label{sec:obs}

The 860 $\mu$m ALMA dust polarization observations (2017.1.00834.S, PI: Adriana Pohl) were carried out on 2018 August 19.
The antenna configurations were  C43-2 with 45 antennas.
Four spectral windows (spws) were set in both the lower (2 spws) and upper (2 spws) sidebands, with 64 channels per spw and 31.25 MHz channel width, providing a bandwidth of $\sim7.5$ GHz in total.
The central frequencies in spws are 349.7 GHz, 351.5 GHz, 361.6 GHz, and 363.5 GHz, respectively.
The bandpass and the gain calibrations were performed by observations of J1427-4206 and J1625-2527,
respectively, and the polarization calibration was performed by
observations of J1427-4206.
The polarization calibrator was observed $3-4$ times with $\sim8$ min integration time during each execution in order to calibrate the instrumental polarization ({\it D}-terms), cross-hand delay, and cross-hand phase.
The total integration time for the target was about 90 min.
The reduction and calibration of the data were done with CASA version 5.1.1 \citep{mcm07} in a standard manner.
A detailed description of the data reduction is given in \citet{nag16}.

All images (Stokes {\it I}, {\it Q}, and {\it U}) were reconstructed with the CASA task {\it tclean} with Briggs weighting with a robust parameter of 0.5.
The beam size of the final product is $0\farcs49\times0\farcs39$, corresponding to a spatial resolution of $\sim59\times47$ au at the assumed distance of 120 pc.

Stokes {\it Q} and {\it U} components produce polarized intensity ($PI$). Note that we ignore the Stokes {\it V} component in this study because it has not been well characterized for ALMA. 
The $PI$ value has a positive bias because it is always a positive quantity. This bias has a particularly significant effect in low-signal-to-noise measurements. We thus debiased the polarized intensity map as
$PI=\sqrt{Q^2+U^2-\sigma_{\rm PI}^2}$.
The root-mean-square (rms) noise of the Stokes {\it I}, {\it Q}, and {\it U}, and the polarized intensity was derived to be $\sigma_{\it I}=3.3\times10$ $\mu$Jy beam$^{-1}$, $\sigma_{\it Q}=3.0\times10$ $\mu$Jy beam$^{-1}$, $\sigma_{\it U}=3.3\times10$ $\mu$Jy beam$^{-1}$, and $\sigma_{\rm PI}=4.4\times10$ $\mu$Jy beam$^{-1}$, respectively. 
The polarization fraction ($P_{\rm frac}=PI/I$) was derived only where detection was above the threshold 3$\sigma_{\rm PI}$.
The 1$\sigma$ error of the polarization angle and polarization fraction is derived to be $\sim1^\circ$ and $\sim0.1$\%, respectively, on average.

 We also utilize the VLA data taken by  \citet{van15}.
The observations were carried out on 2015 January - February in the CnB and B configurations.
The observing wavelength is $\lambda=8.8$ mm (34 GHz).
The beam size of the VLA observations is $0\farcs46\times0\farcs26$ with a position angle of $21^\circ$.
The detailed comparison of the ALMA 440 $\mu$m and the VLA 8.8 mm data are discussed by \citet{van15}.

\section{Results}\label{sec:res}

\subsection{The proper motion of IRS 48}

Before we show the results of the ALMA polarization data, we compare the continuum peak positions identified by the ALMA 860 $\mu$m (this study) and previous VLA 8.8 mm observations.
\citet{van15} found the same peak position between the ALMA 440 $\mu$m and VLA 8.8 mm continuum emission because these observations were carried out  within 6 months.
In contrast, since the period in between our and previous observations is  3.5 yr, the proper motion may be not negligible.

\begin{figure*}[htbp]
  \begin{center}
  \includegraphics[width=16.cm,bb=0 0 2950 1229]{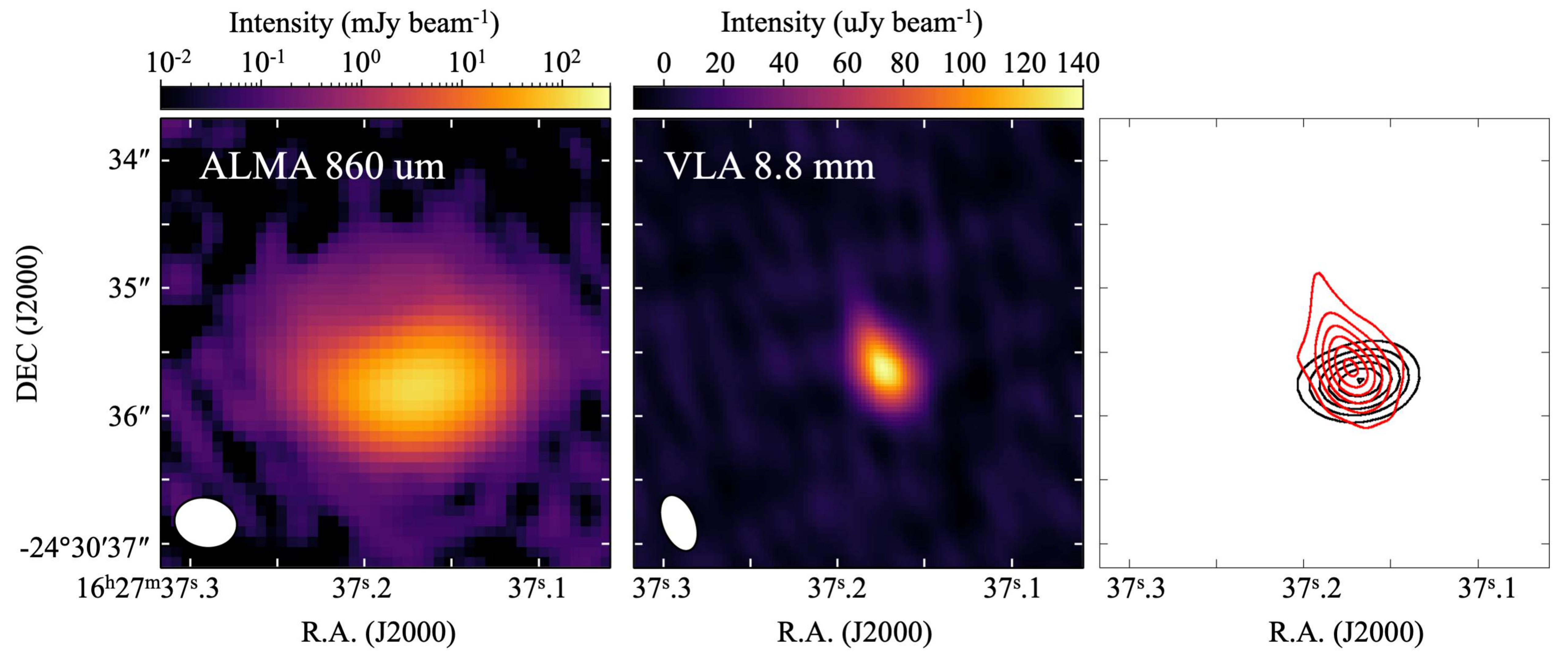}
  \end{center}
  \caption{ALMA  and VLA  observations of dust continuum emission at 860 $\mu$m and 8.8 mm toward the Oph IRS 48 disk.
  (Left) The ALMA 860 $\mu$m image (logarithmical color scale); (middle) the VLA 8.8 mm image (linear color scale); (right) the overlay of the VLA8.8 mm contours in red (taken at 0.03, 0.048, 0.066, 0.084, 0.102, and 0.12 Jy beam$^{-1}$) on the ALMA 860 $\mu$m contours in black (taken at 12, 36, 60, 84,108, and 132 $\mu$Jy beam$^{-1}$).  }
  \label{alma_vla}
\end{figure*}

Figure \ref{alma_vla} shows images of the dust continuum emission at the ALMA 860 $\mu$m and VLA 8.8 mm observations.
The detailed structures are described in the following section. Here, we point out that an offset of the peak positions between ALMA and VLA observations is found.
The ALMA 860 $\mu$m continuum peak is slightly shifted to South-West direction.
The offset is $\Delta {\rm R.A.} \sim -0.00564{\rm s}(=0\farcs0846)$ and $\Delta {\rm DEC} \sim-0\farcs114$, respectively.
Even though the offset of $\sim0\farcs1$ is smaller than the beam size, we suggest that IRS 48 has a proper motion of ($-24.2\pm5.7,-32.5\pm5.7$) mas yr$^{-1}$ because the ALMA positional accuracy is $\sim10-20$ mas according to the ALMA Technical Handbook.
The presented images in below are corrected for the proper motion.
However, the proper motion of IRS 48 is estimated to be ($-9,-24$) mas yr$^{-1}$ by  \citet{gai18}. The different measurements of the proper motion might indicate an orbital motion of the disk. Further observations are needed to investigate the proper motion of the protostar and the orbital motion of the disk.

\begin{figure*}[htbp]
  \begin{center}
  \includegraphics[width=16.cm,bb=0 0 1500 1685]{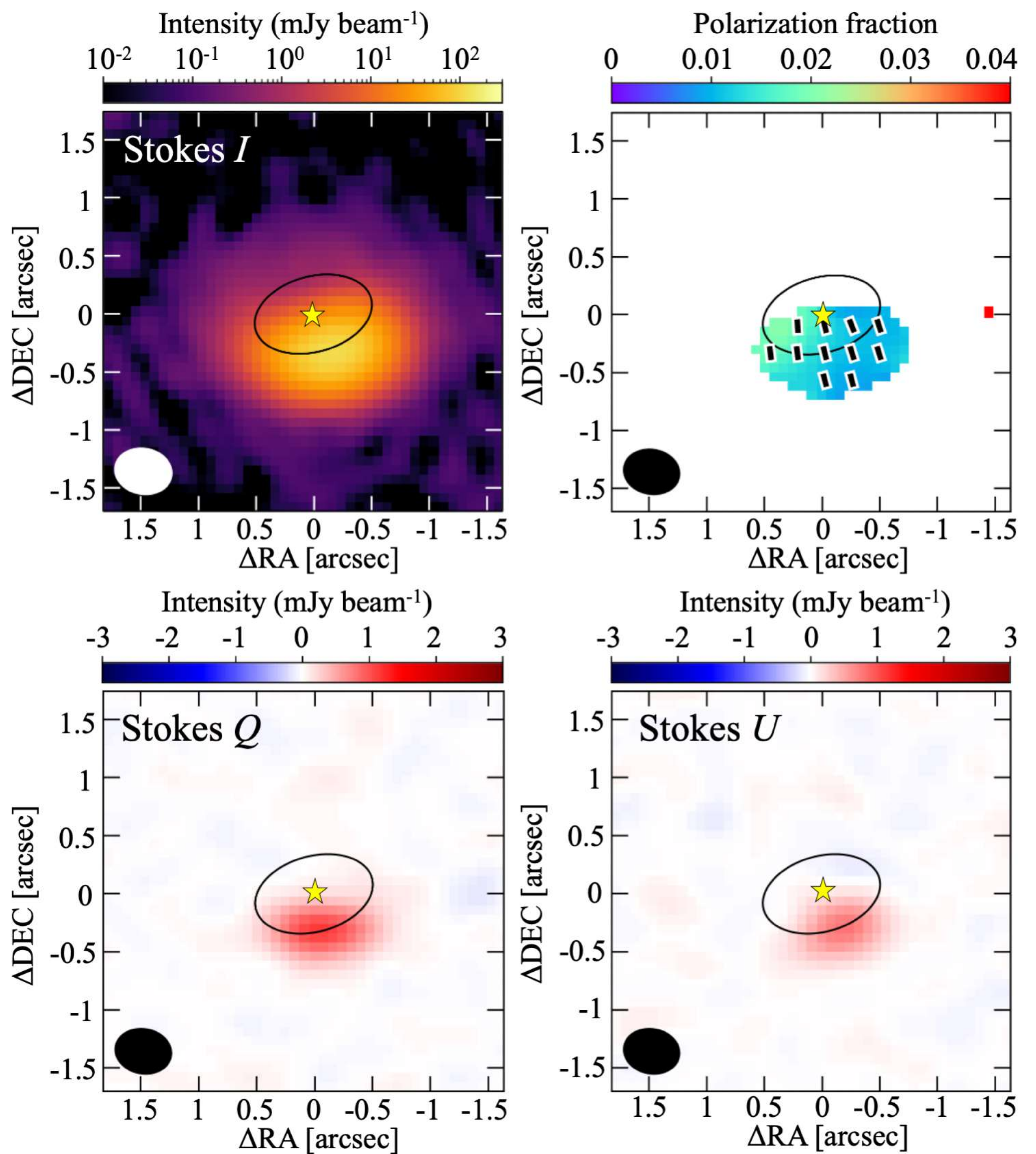}
  \end{center}
  \caption{860 $\mu$m dust continuum polarization in IRS 48. The upper panels show the total intensity (Stokes {\it I}, left) and polarization fraction and position angle (right). The lower panels show the Stokes {\it Q} (left) and {\it U} (right) emission. The black circles indicate the position of the 60 au ring. The polarization vectors and the polarization fraction are shown where the polarized intensity is higher than 3$\sigma_{\rm PI}$. Note that the lengths of the polarization vectors are set to be the same.  The stellar position is indicated by the star marker and is set to 16$^h$27$^m$37$^s$.2 -24$^{\circ}$30'35$\farcs$5.
  }
  \label{obs}
\end{figure*}

\subsection{ALMA Polarization Data}
The results of the ALMA polarization data are shown in Figure \ref{obs}.
The black line represents the radius of the dust peak in the disk at 60 au identified by \citet{van13}.
Stokes {\it I} image shows an oval structure rather than the lopsided disk because the beam size is insufficient to resolve the substructure previously reported in \citet{van13}.
The peak emission is 120 mJy beam$^{-1}$ toward the southeast of the disk, corresponding to 13 K in brightness temperature.
The integrated flux of the continuum emission is 190 mJy.
Stokes {\it Q} and {\it U} emissions is detected toward the peak emission of the Stokes {\it I} with intensities of $\sim1$ mJy beam$^{-1}$.

The polarization fraction, defined as $P_{\rm frac} = {\rm POLI}/{\it I}$, is on average $\sim1-2$\%, and the polarization vectors show an angle of $\sim10^\circ$, which is similar to the orientation of the disk minor axis because the position angle of the disk (i.e., its major axis) is derived to be $\sim100^\circ$ \citep{gee07,bru14}.

\subsection{Stokes {\it I} Continuum Data}

The polarization observations were carried out with high sensitivity.
Therefore, we are able to detect faint emission from the continuum.
The previous observations of ALMA dust continuum emission were not able to measure the brightness contrast between the north and south regions because the north side of the ring was not detected.

\begin{figure}[htbp]
  \includegraphics[width=8.cm,bb=0 0 2224 1619]{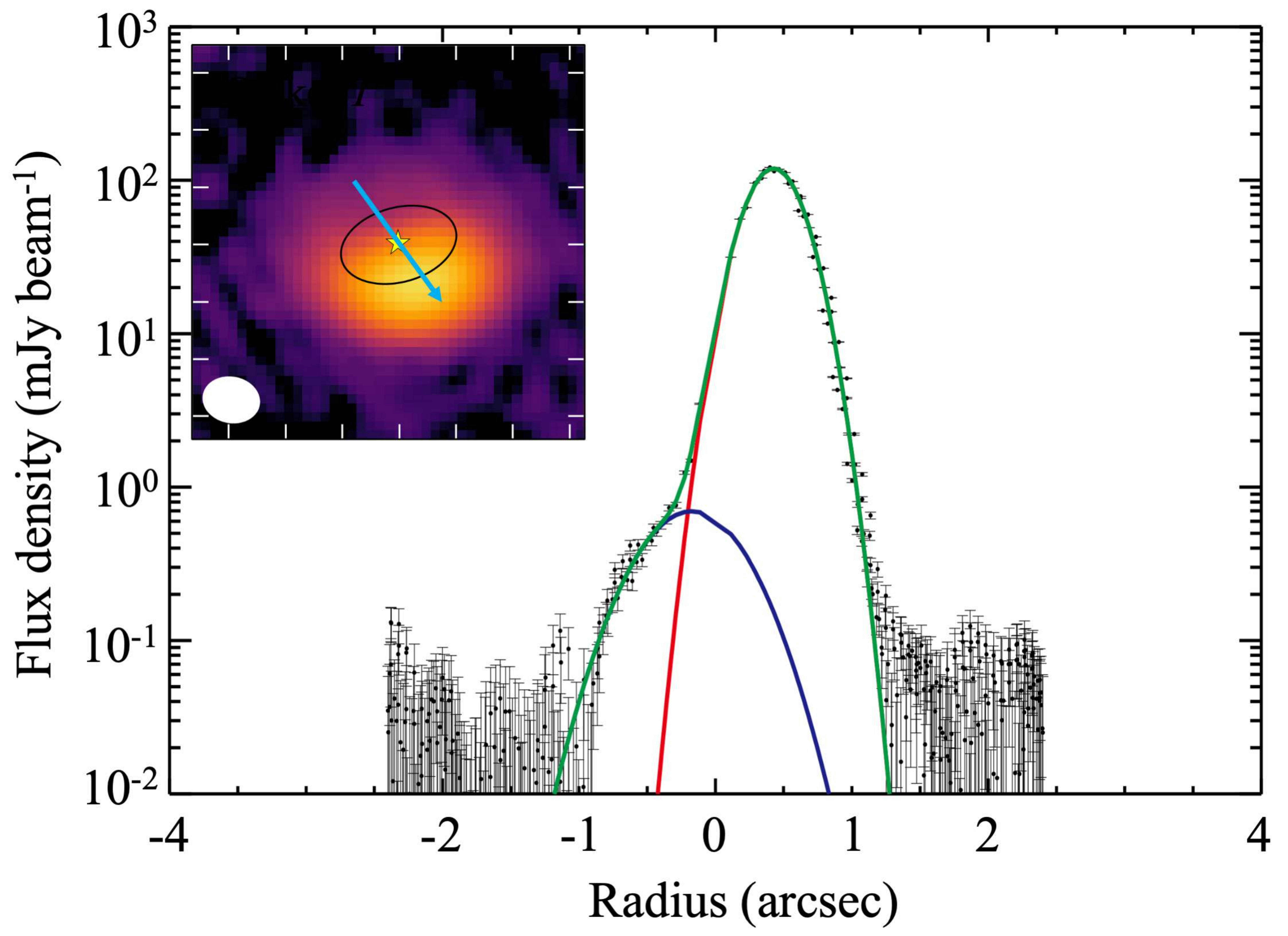}
\caption{The radial profile of the continuum emission (Stokes {\it I}) along the position angle of ${\rm P.A.}=36\pm15^{\circ}$ including the peak intensity. The origin is set to be the central star. The green line is the two-component Gaussian fitting. The red line is the peak component of the southern part of the disk, while the blue line indicates the detection of the continuum emission from the northern side of the ring. The intensity contrast in the flux is 170.}
\label{radial_plot}
\end{figure}

We plot in Figure \ref{radial_plot} the intensity profile of the dust continuum emission along  the position angle of ${\rm P.A.}=36\pm15^{\circ}$ to include the peak intensity.
 We note that the peak position does not coincide with the disk minor axis.
The peak intensity is found at a distance of 0$\farcs$4 from the central star toward the southern direction with the intensity of 120 mJy beam$^{-1}$.
In addition to the peak emission, we find the second component on the opposite side of the disk toward the northern direction.
We fit the profile with a two component Gaussian profile, which is shown in Figure \ref{radial_plot}.
The red line shows the component of the main Gaussian profile, and the blue line shows the second component. The green line indicates the combination of these two components.
With the two component Gaussian fitting, the peak intensity ($I_{\rm peak}$), the center position ($r_{\rm peak}$), and the Gaussian width ($\sigma_{\rm peak}$) are derived to be $I_{\rm peak}=120$ Jy beam$^{-1}$, $r_{\rm peak}=0\farcs4$, and $\sigma_{\rm peak}=0\farcs2$ for the main component and $I_{\rm peak}=0.70$ mJy beam$^{-1}$, $r_{\rm peak}=0\farcs2$, and  $\sigma_{\rm peak}=0\farcs3$ for the second component, respectively.

The brightness contrast between the southern and northern sides of the disk is $\sim170$ (without the deconvolution with the beam size), which is one order or two orders of magnitudes stronger contrast than that found in other lopsided disks such as HD 142527, LkH$\alpha$ 330, and HD135344B \citep{cas13,fuk13,ise13,caz18}.
The center position of the second component is slightly closer to the central star compared with the main component. By taking into account that \citet{van16} revealed that the gas cavity is two times smaller than the dust cavity, these results may suggest that the northern part of the disk only contains smaller dust grains colocated with the gas.
This small grain distribution was found by \citet{gee07,van13}.
Such different dust size distributions are also suggested in HD 142527 by the ALMA polarization observations \citep{oha18}.
However, the beam smearing from the southern part may also affect the northern emission. Therefore, further observations with higher angular resolution will allow us to reveal differences in the center position and width of the north-south ring structure.

\subsection{ VLA Continuum Data}
\begin{figure}[htbp]
  \includegraphics[width=8.cm,bb=0 0 746 827]{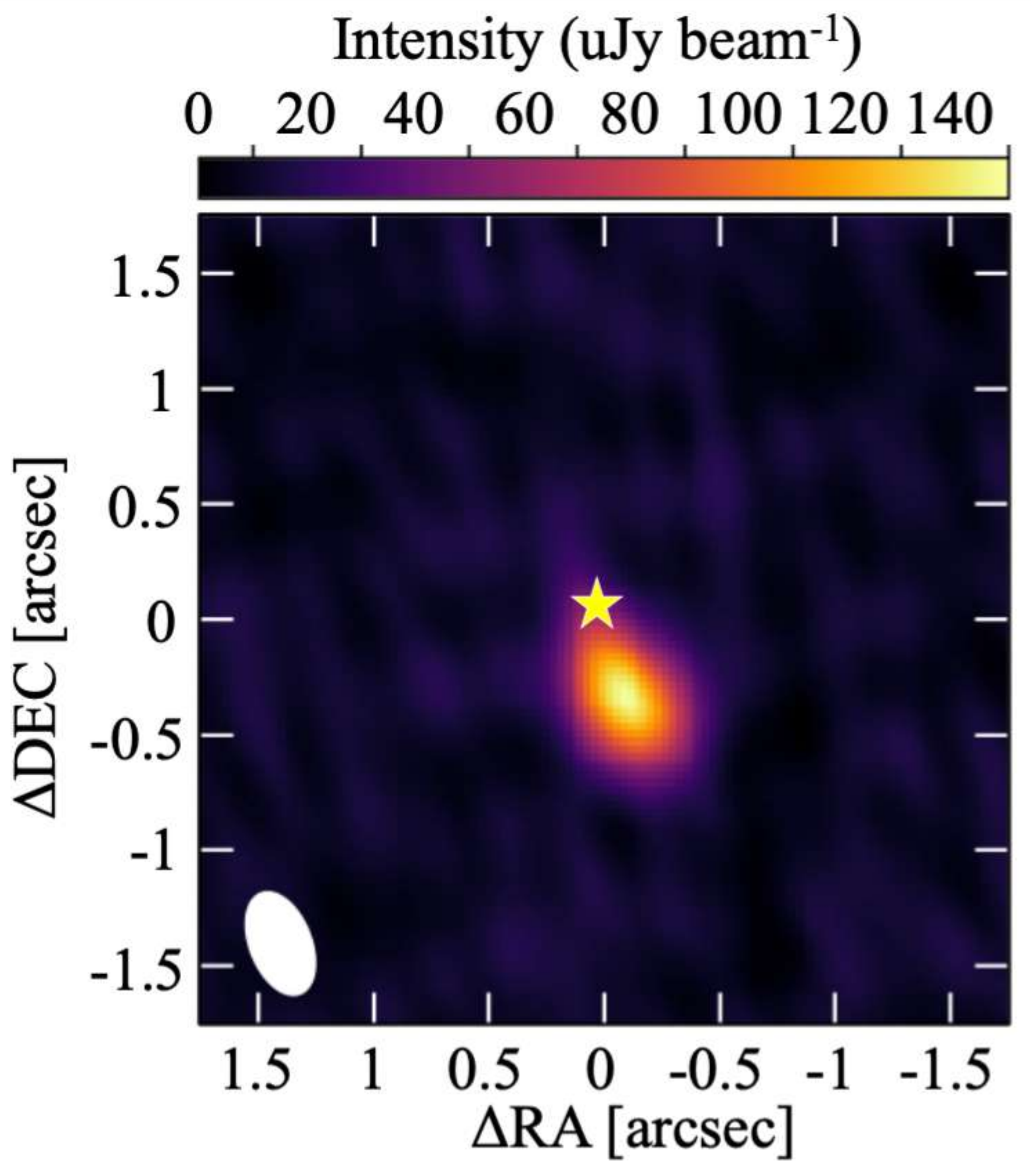}
\caption{The VLA observations of dust continuum emission at 8.8 mm of the IRS 48 disk. 
The stellar position was set to 16$^h$27$^m$37$^s$.2 -24$^{\circ}$30'35$\farcs$3.
}
\label{vla}
\end{figure}

Here, we show the VLA image in Figure \ref{vla} again because we analyze the grain size by using both ALMA polarization and VLA continuum data (see Section \ref{sec:model}). 
The image is marginally resolved.
Continuum emission is detected in the dust trap with the peak intensity of 138 $\mu$Jy beam$^{-1}$. 
By comparing with the ALMA 440 $\mu$m image, the VLA image is found to be more concentrated in the dust trap, suggesting that the VLA continuum emission traces larger dust grains.
Note that \citet{van15} reported a small point-like contribution from an unresolved inner disk and/or free-free/synchrotron emission from ionized gas close to the star.
However, the derived flux of that inner component is only 36 $\mu$Jy, which is only a small fraction of the total flux of 250 $\mu$Jy.
Therefore, we assume that the most of the VLA emission comes from dust grains in this study.

\section{Radiative Transfer Modeling}\label{sec:model}
The morphology of the polarization vectors parallel to the disk minor axis is similar to those found in other inclined protoplanetary disks \citep{ste14,kat17,ste17,lee18,cox18,gir18,sad18,hul18,har18,bac18,den19,har19,mor19}, and is consistent with the self-scattering model \citep{kat15,yang16}.
The polarization fraction of $1-2$\% is also expected by self-scattering.
Therefore, the 860 $\mu$m dust polarization of the IRS 48 disk is most likely caused by self-scattering, which indicates a maximum grain size of $\sim100$ $\mu$m because self-scattering becomes the most efficient when the grain size is $\sim\lambda/2\pi$.

On the other hand, \citet{van15} detected 34 GHz ($\lambda\sim8.8$ mm) dust continuum emission with VLA as well as 680 GHz dust continuum emission with ALMA. The VLA image suggests that centimeter-sized grains exist and appear to be more concentrated in azimuth than the millimeter-sized grains as predicted in analytical dust models of azimuthal pressure maxima \citep{bir13}.
Therefore, the ALMA and VLA data imply different maximum grain sizes.
Here, we discuss the possible scenarios to explain those different measurements from the ALMA polarization and VLA continuum emission.

\begin{table}
\centering
\caption{\label{table} Models of possible dust distributions for radiative transfer calculations}
\begin{tabular}{l}\hline\hline
Optically thin model  \\\hline
1. Optically thin with $a_{\rm max}\sim100$ micron  	\\
2. Optically thin with $a_{\rm max}\sim100$ micron and $\sim$ cm	\\\hline\hline
Optically thick model  \\\hline
3. Optically thick with $a_{\rm max}\sim100$ micron \\\hline
\end{tabular}
\label{table}
\end{table}

Below, we consider an optically thin case (Section 4.1). 
The maximum grain size is considered to be $\sim100$ $\mu$m for reproducing the ALMA polarization data. Then, we investigate two populations of dust grains with maximum grain sizes of $\sim100$ $\mu$m and $\sim$ cm to take into account the VLA data. 
Next, we consider an optically thick case with dust grains having a maximum grain size of $\sim100$ $\mu$m (Section 4.2).
We summarize the possible dust distributions in Table \ref{table}.

To reproduce the ALMA polarization and VLA continuum emission, we performed radiative transfer calculations with RADMC-3D\footnote{RADMC-3D is an open code of radiative transfer calculations developed by Cornelis Dullemond. The code is available online at: \url{http://www.ita.uniheidelberg.de/~dullemond/software/radmc-3d/}} \citep{dul12} that take into account multiple scattering, as done by \citet{kat15}.
We constructed the continuum image with a two-dimensional fourth power Gaussian intensity profile $I_\nu (r, \phi)$, following \citet{van13,van15}:
\begin{equation}
    I_{\nu}(r,\phi)=I_{\rm c}\exp\Big(\frac{-(r-r_{\rm c})^4}{2r_{\rm w}^4}\Big)\Big(\frac{-(\phi-\phi_{\rm c})^4}{2\phi_{\rm w}^4}\Big),
\label{in}
\end{equation}
\begin{equation}
    I_c=B_{\nu}(T_{\rm d})(1-e^{-\tau_{\rm c}}),
\label{intensity}
\end{equation}
where $r$ is a radius from the central star, $\phi$ is an azimuthal angle, $\tau$ is the optical depth and $B_{\nu}(T_{\rm d})$ is the Planck function.
 Note that $r_{\rm c}, r_{\rm w}, \phi_{\rm c}$ and $\phi_{\rm w}$ denote the peak radius, the radial width of the distribution, the peak azimuthal angle, and the azimuthal width of the distribution, respectively.
The optical depth $\tau_{\rm abs}$ is calculated as $\tau_{\rm abs}=\kappa_{\rm abs}\Sigma_{\rm d}$, where $\kappa_{\rm abs}$ is the absorption opacity and $\Sigma_{\rm d}$ is the dust surface density.

The dust opacity is calculated following \citet{kat16}.
The dust grains are assumed to be spherical and to have a power-law size distribution with an exponent of $q=-3.5$ \citep{mat77} and maximum grain size $a_{\rm max}$.
These assumption are reasonable, as \citet{taz19} suggest that the compact structure of dust grains  is better to explain the observed polarization fraction than the fluffy dust grains with high fractal dimension.
This maximum grain size is considered to be the representative grain size in the following discussion. The opacity was calculated using Mie theory. 
The composition was assumed to be a mixture of silicate (50\%) and water ice (50\%) \citep{pol94,kat14}. We used the refractive index of astronomical silicate \citep{wei01} and water ice \citep{war84} and calculated the absorption and scattering opacity based on effective medium theory using the Maxwell-Garnett rule \cite[e.g.,][]{boh83,miy93}. 
 For example, the absorption opacity of dust grains with $a_{\rm max}=140$ $\mu$m is calculated to be 0.88 g$^{-1}$ cm$^{2}$ at 860 $\mu$m wavelength and 0.011 g$^{-1}$ cm$^{2}$ at 8.8 mm wavelength.

Because the absorption opacity, $\kappa_{\rm abs}$, varies depending on dust grain size, we change the surface density to match the observed intensities as follows:
\begin{equation}
    \Sigma_{\rm d}(r,\phi)=\Sigma_{\rm 0}\exp\Big(\frac{-(r-r_{\rm c})^4}{2r_{\rm w}^4}\Big)\Big(\frac{-(\phi-\phi_{\rm c})^4}{2\phi_{\rm w}^4}\Big),
\label{sigma}
\end{equation}
 where $\Sigma_{\rm 0}$ is the peak surface density.

The disk inclination is set to be $50^\circ$ with assuming that the southern region is the near side \citep{bru14}.
The position angle is assumed to be $110^\circ$ to reproduce the observations.

By using the image of 440 $\mu$m continuum emission, \citet{van15} found the best fit for $r_{\rm c}=61$ AU, $r_{\rm w}=14$ AU, $\phi_{\rm c}=100^\circ$, and $\phi_{\rm w}=41^\circ$.
We use these parameters for our calculations for $\sim100$ $\mu$m dust grains.

\citet{oha19} found that the dust scale height is also a key parameter to induce the polarization by the self-scattering. Therefore, we set an additional dust settling parameter $f_{\rm set}$ such that $h_{\rm d}=h_{\rm g}/f_{\rm set}$, where $h_{\rm g}$ is the gas scale height, to mimic grain settling.
The gas scale height is written as $h_{\rm g}=c_{\rm s}/\Omega_{\rm K}$, where $c_{\rm s}$ is the sound speed and $\Omega_{\rm K}$ is the Keplerian angular velocity.
The vertical density distribution is assumed to be Gaussian with a dust scale height $h_{\rm d}$ such that $\rho_{\rm d}=\Sigma_{\rm d}/(\sqrt{2\pi}h_{\rm d})\exp(-z^2/h_{\rm d}^2)$.

\subsection{An Optically Thin Case}

We assume a dust temperature of 60 K at 60 au \citep{bru14}. Then, for low optical depths, the temperature profile is assumed to have a power law index of $-0.5$ \citep{ken87} as
\begin{equation}
T_{\rm d}= 60\ {\rm K}\ \Big(\frac{r}{60\ {\rm au}}\Big)^{-0.5}. 
\label{temp}
\end{equation}

The brightness temperature of the 440 $\mu$m wavelength dust continuum emission  was derived to be 31 K at 60 au with ALMA high resolution observations \citep{van15}.
Therefore, the temperature of 60 K, higher than the brightness temperature, suggests that the continuum emission even at 440 $\mu$m is marginally optically thick, as long as significant beam dilution does not occur.

\subsubsection{Single Grain Population with Maximum Grain Size $a_{\rm max}\sim100$ $\mu$m}\label{single}
As a simple case (Model 1 in Table \ref{table}), we consider a grain size population with a single power-law size distribution with $n(a)\propto a^{-3.5}$ and $a_{\rm min}=0.1$ $\mu$m.
The maximum grain size $a_{\rm max}$ is the main parameter. We, here, use $a_{\rm max}=140$ $\mu$m because the dust grains with $a_{\rm max}=140$ $\mu$m produce the polarization with the maximum efficiency at 860 $\mu$m.

The dust surface density is set to be $\Sigma_{\rm 0}=0.26$ g cm$^{-2}$ at 61 au in equation (\ref{sigma}) to recover the flux of the ALMA 860 $\mu$m continuum emission.
The corresponding optical depth is $\tau_{\rm abs}=0.23$ because the dust absorption opacity is calculated to be 0.88 g cm$^{-2}$ from Mie theory by assuming a maximum grain size of 140 $\mu$m.

We use the dust settling parameter $f_{\rm set}=10$, which is the case that dust grains are well settled into the midplane of the disk, since other disks such as HL Tau are suggested to have the low dust scale height \citep{pin16}. We discuss the dust scale height later in Section \ref{dust_scale}.

\begin{figure*}[htbp]
  \includegraphics[width=18.cm,bb=0 0 2558 1607]{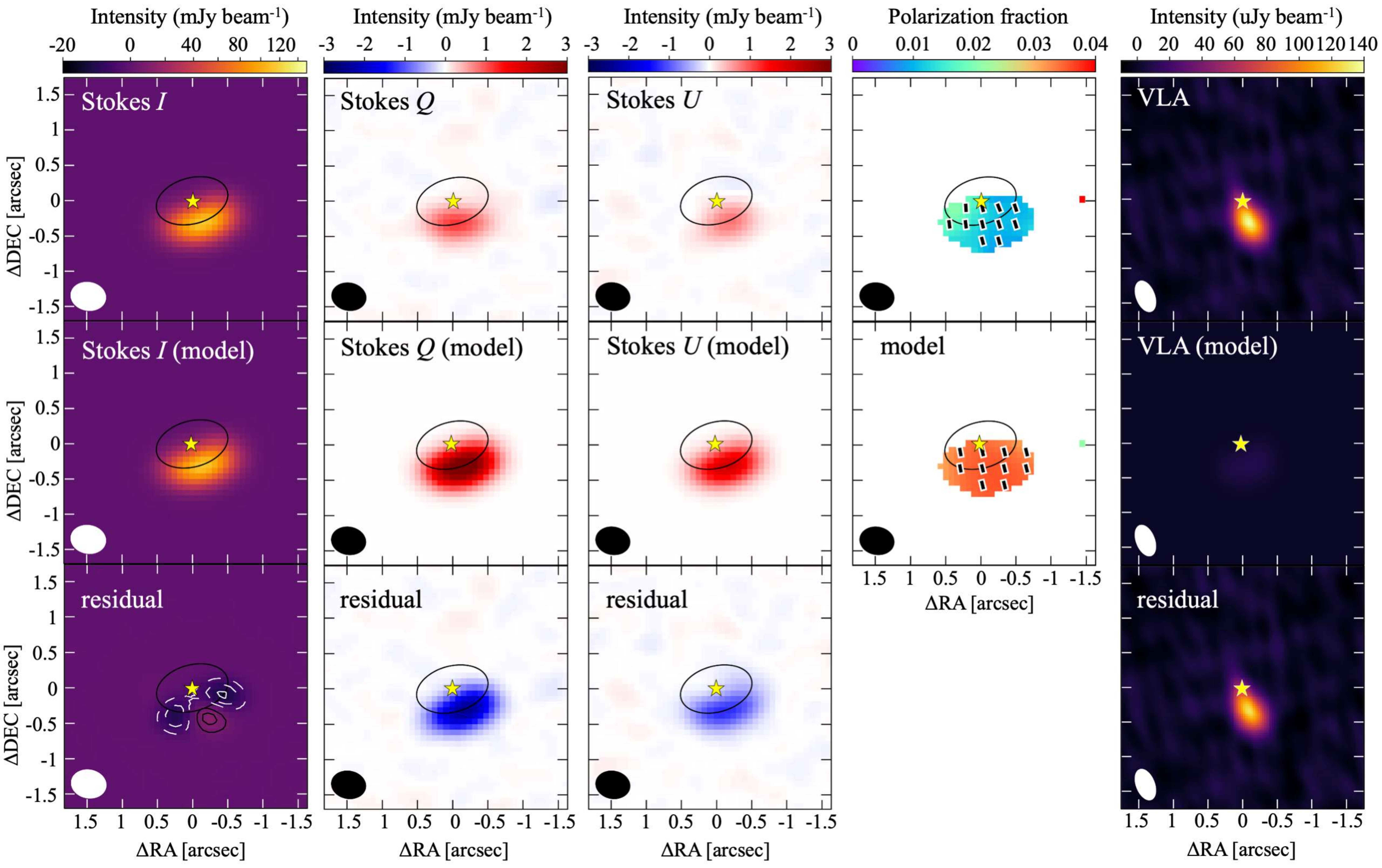}
\caption{(Top row) The observation images of the ALMA 860 $\mu$m dust continuum polarization and VLA 8.8 mm dust continuum data. (Middle row) The model images of the radiative transfer calculations for the total intensity (Stokes {\it I}), Stokes {\it Q}, and {\it U} emission, polarization fraction overlaid with polarization vectors, and the VLA 8.8 mm dust continuum. The model assumes optically thin emission with the single grain population with a maximum grain size of 140 $\mu$m . The dust settling parameter is set to $f_{\rm set}=10$. (Bottom row) The residual images  in the same colorscale as the data image. The black solid and white dashed contours in the Stokes {\it I} residual map indicate the positive and negative intensity levels with $\pm5.0$ mJy beam$^{-1}$ intervals.
}
\label{test1}
\end{figure*}

Figure \ref{test1} shows the comparisons of ALMA Band 7 polarization and VLA continuum observations and the radiative transfer calculations.
The calculations are performed with the same model at the two observing wavelengths of $\lambda=860\mu$m and 8.8 mm. 
The images are convolved with the beam sizes of $0\farcs493\times0\farcs392$ for ALMA and $0\farcs46\times0\farcs26$  for VLA, respectively.

The ALMA Stokes images and our model are similar.
The Stokes {\it Q} and {\it U} emission have their peaks at the continuum peak position.
The polarization vectors seem parallel to the disk minor axis even though a slight offset from the observations are found. The polarization vectors parallel to the disk minor axis are consistent with the self-scattering model.
The small differences from the observations may be caused by the radiation fields we assume. The dust scale height and/or the intensity profile of our model may be slightly different from the IRS 48 disk.
Since the polarization vector depends on the anisotropy of the radiation field, the polarization vector varies with the intensity profile and the dust scale height. We discuss the polarization variations by changing the dust scale height in Section \ref{dust_scale}.
The polarization fraction is derived to be $\sim3.5$\%, which is slightly higher than the observed values. To fit the observed polarization fraction (mostly 1-2\%), the grain size would need to be slightly smaller or larger than 140 $\mu$m ($a_{\rm max}\sim60 - 180$ $\mu$m), since the 140 $\mu$m dust grain is the most efficient to produce the polarization due to the scattering, and thus yields the highest value of polarization fraction.
Another way to fit the observed polarization fraction is to change the dust compositions or structures. The polarization fraction can be lower if the porosity of the dust grains are higher \citep{taz19}.

The observations show a maximum polarization fraction of $\sim2\pm0.5$\% toward the eastern side of the continuum peak, whereas the western side of the disk has a lower polarization fraction of $\sim1\pm0.1$\%.
This may suggest different grain size distributions along the azimuthal direction.
However, the difference in grain size should be small because the dust size needs to be $\sim30-300$ $\mu$m to produce any scattering polarization at 860 $\mu$m  (grains larger or smaller than those values do not produce the scattering-induced polarization at 860 $\mu$m wavelength). 
Thus, we suggest that pure self-scattering with a maximum grain size of $\sim100$ $\mu$m can explain the ALMA polarization observations.

However, Figure \ref{test1} also shows the comparison between the VLA observations and model.
The model shows a significant discrepancy from the observations, with a peak intensity of $\sim10$ $\mu$Jy, which is 10 times lower than the VLA observation because dust grains as small as 140 $\mu$m are insufficient to produce the emission at 8.8 mm wavelength.
Therefore, this model with a single population of dust grains with $a_{\rm max}\sim100$ $\mu$m cannot explain the VLA observations.

\subsubsection{Two Populations of Dust Grains with the Maximum Grain Sizes of $a_{\rm max}\sim100$ $\mu$m and $a_{\rm max}\sim$ cm}
In Section \ref{single}, we discuss dust grains with a maximum grain size of $140$ $\mu$m.
We found that the 140 $\mu$m dust grains are not large enough to explain the VLA observations even though the model can match the ALMA polarization observations.

\begin{figure*}[htbp]
  \includegraphics[width=18.cm,bb=0 0 2564 1607]{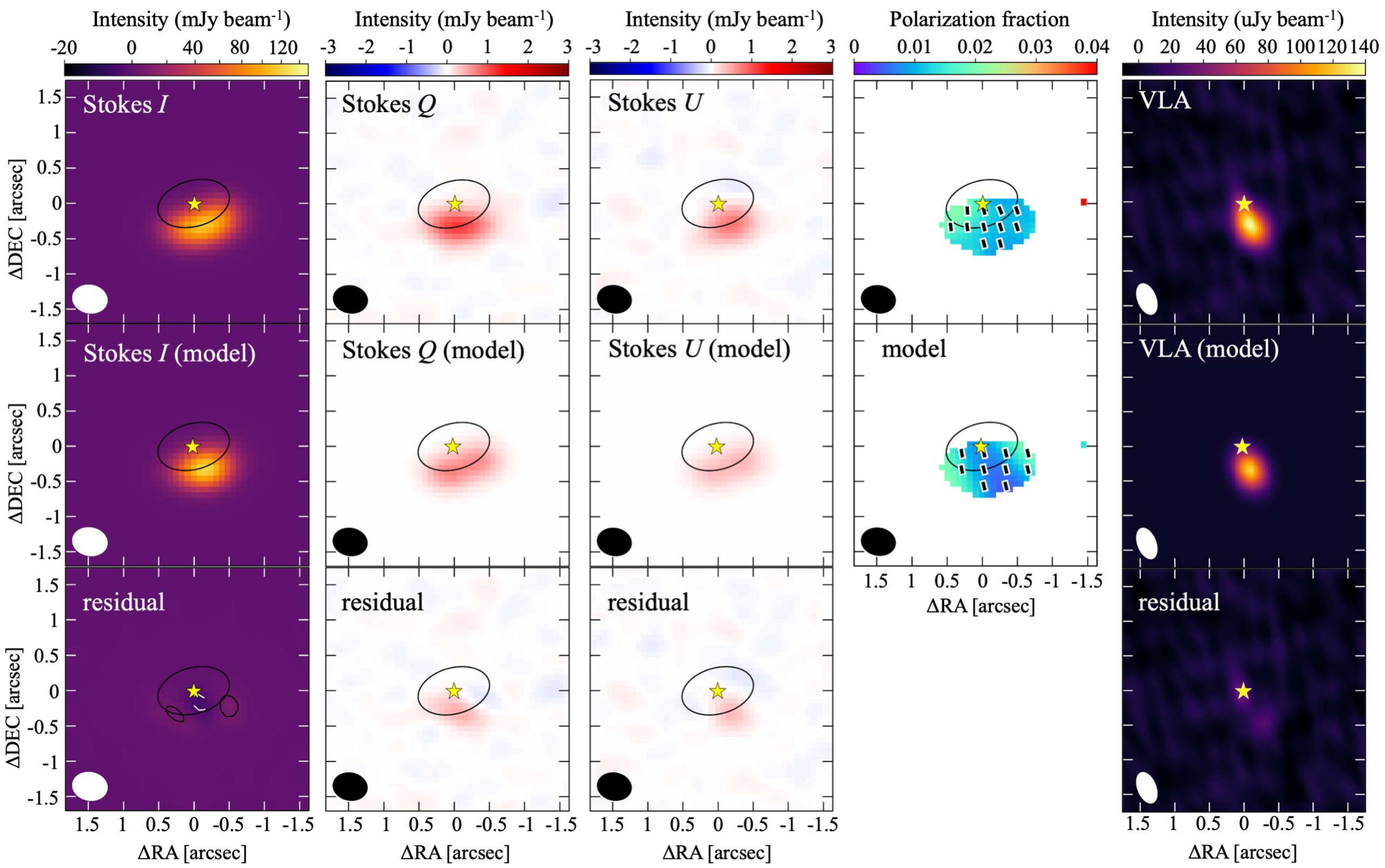}
\caption{ Same as Figure \ref{test1} but assuming two populations of dust grains. The 5 cm dust grains are located at the continuum peak with an azimuthal width of $\phi_{\rm w}=18^\circ$. The 140 $\mu$m dust grains are distributed in a larger crescent with the azimuthal width of $\phi_{\rm w}=41^\circ$.}
\label{test2}
\end{figure*}

Here, we discuss two populations of dust grains to explain both observations of ALMA polarization and VLA continuum emission under the optically thin conditions (Model 2 in Table \ref{table}). 
We assume that one consists of a maximum grain size of 140 $\mu$m and the other consists of a maximum grain size of $\sim1$ cm.
The 140 $\mu$m dust grains represent the ALMA polarization emission, while the $\sim1$ cm dust grains represent the VLA continuum emission.
Here we use a maximum grain size of 5 cm with a power-law size distribution of $q=-3.5$ to reproduce the VLA continuum emission. 
It should be noted that the spectral index, $\beta$, depends not only on the grain sizes of the emitting dust, but also on other factors such as dust chemical composition, porosity, geometry, as well as on the grain size distribution \citep{tes14}.
Therefore, we need to investigate the power-law of size distribution and dust compositions for reproducing the lower spectral index in future studies.

According to \citet{van15}, we consider different spatial distributions for these two populations.
For the 5 cm dust grains, we assume that the azimuthal width of the crescent is $\phi_{\rm w}=18^\circ$ in equation (\ref{in}) in order to reproduce the VLA continuum emission. 
The 140 $\mu$m dust grains are assumed to have a larger azimuthal crescent with an azimuthal width of $\phi_{\rm w}=41^\circ$. The 140 $\mu$m dust grains set outside of the region where the 5 cm dust grains are distributed to reproduce the images of the ALMA 440 $\mu$m continuum emission and 860 $\mu$m dust polarization emission.
We remove the 140 $\mu$m dust grains in the dust trap as traced by the centimeter emission because dust grains are considered to have already grown to 5 cm in this region.

To reproduce the intensities of ALMA 440 $\mu$m, ALMA 860 $\mu$m, and VLA 8.8 mm dust continuum emission, the surface density is set to be $\Sigma_0=0.65$ g cm$^{-2}$ for the 5 cm dust grains.
The surface density of the  140 $\mu$m dust grains is set to be $\Sigma_0=0.1$ g cm$^{-2}$ but is excluded in the center of the dust trap (no 140 $\mu$m dust grains are colocated with the 5 cm dust grains). 
These dust distributions keep the intensity profile revealed by ALMA 440 $\mu$m dust continuum observations.
The continuum image of this model for the ALMA 440 $\mu$m observations is discussed in Section \ref{two_populations}.
The dust settling parameter of $f_{\rm set}=10$ is used.

Figure \ref{test2} compares the images of ALMA Band 7 polarization and our model.
Similar to Figures \ref{obs} and \ref{test1}, the polarization vectors are parallel to the disk minor axis. The polarization fraction shows $\sim0.5-2.0$\% and it decreases to $\sim0.5$\% at the continuum peak.
This is because only the 5 cm dust grains are located at this region, which do not produce the scattering-induced polarization.
Our large beam size cannot spatially resolve the different distributions of the two dust grains. Therefore, the polarization appears even in the continuum peak region.

By comparing the Stokes images of the observations and models, we find that the model images of the Stokes {\it Q} and {\it U} emission show slight drops at the continuum peak and thus have double-peaked distributions even though the observations only show single peaks of polarized emission at the continuum peak.
This is also shown in the residual maps of Stokes {\it Q} and {\it U} images because the emission is still remained in the center of the dust trap.
Furthermore, the polarization fraction of our model is lower than the observations even though we use a maximum grain size of 140 $\mu$m, which is the most efficient to produce the polarization.
Therefore, dust grains with a maximum grain size of $\sim100$ $\mu$m may need to be located also in the continuum peak unlike our model.
Furthermore, such 100 $\mu$m sized dust grains need to dominate the continuum emission to reproduce the polarization fraction as shown in Figure \ref{test1}.
These conditions may be unlikely because the  larger dust grains will be more concentrated on the dust trap \citep{bir13}. Furthermore, it may also be difficult to have two distinct populations of dust grains with sizes of 140 $\mu$m and 5 cm in the same region.
However, we cannot rule this model out based on our spatial resolution.
It may be possible that the 5 cm dust grains are more concentrated with the azimuthal width of $\phi_{\rm w}<18^\circ$.
Further observations with higher spatial resolution can testify whether there are $\sim100$ $\mu$m dust grains at the continuum peak position (dust trap position) that are producing the polarization.

Figure \ref{test2} also shows the comparison of the VLA continuum emission and our model.
We find that our model recovers the continuum emission by including the 5 cm dust grains because the large dust grains efficiently produce emission at longer wavelengths.
The peak intensity of our model shows 120 $\mu$Jy beam $^{-1}$, which is slightly lower than 138 $\mu$Jy beam $^{-1}$ derived by the VLA observations.
However, it may be possible to reproduce the VLA continuum emission by changing the power-law size distribution of $q$ or changing the dust chemical compositions.
It should be noted that this study does not make an optimal fitting, and we give an idea of whether the model can explain the observations.

These results may indicate that the model using two grain populations can explain generally both observations of the ALMA polarization and VLA continuum emission. The 5 cm grains are more concentrated and the 140 $\mu$m dust grains are distributed in a larger area.
These different spatial distributions of dust grains are consistent with the dust trapping scenario.

However, the simple dust trapping scenario may not be the case because the two distinct grain sizes of 140 $\mu$m and 5 cm are used without considering any other maximum grain sizes in our model, for example, a perhaps more realistic broader size distribution. The model assumes that the $\sim100$ $\mu$m dust grains need to be widely distributed outside of the 5 cm dust grains.
If the maximum grain size were to have an azimuthal distribution, the polarization would be produced only where the grain size is $\sim100$ $\mu$m, which would lead to a significant local reduction of the polarization degree due to the beam dilution \citep{poh16}.

\subsubsection{Dust trapping model for ALMA polarization observations}\label{sec:dust_trap}

\begin{figure}[htbp]
  \includegraphics[width=8.cm,bb=0 0 2176 1460]{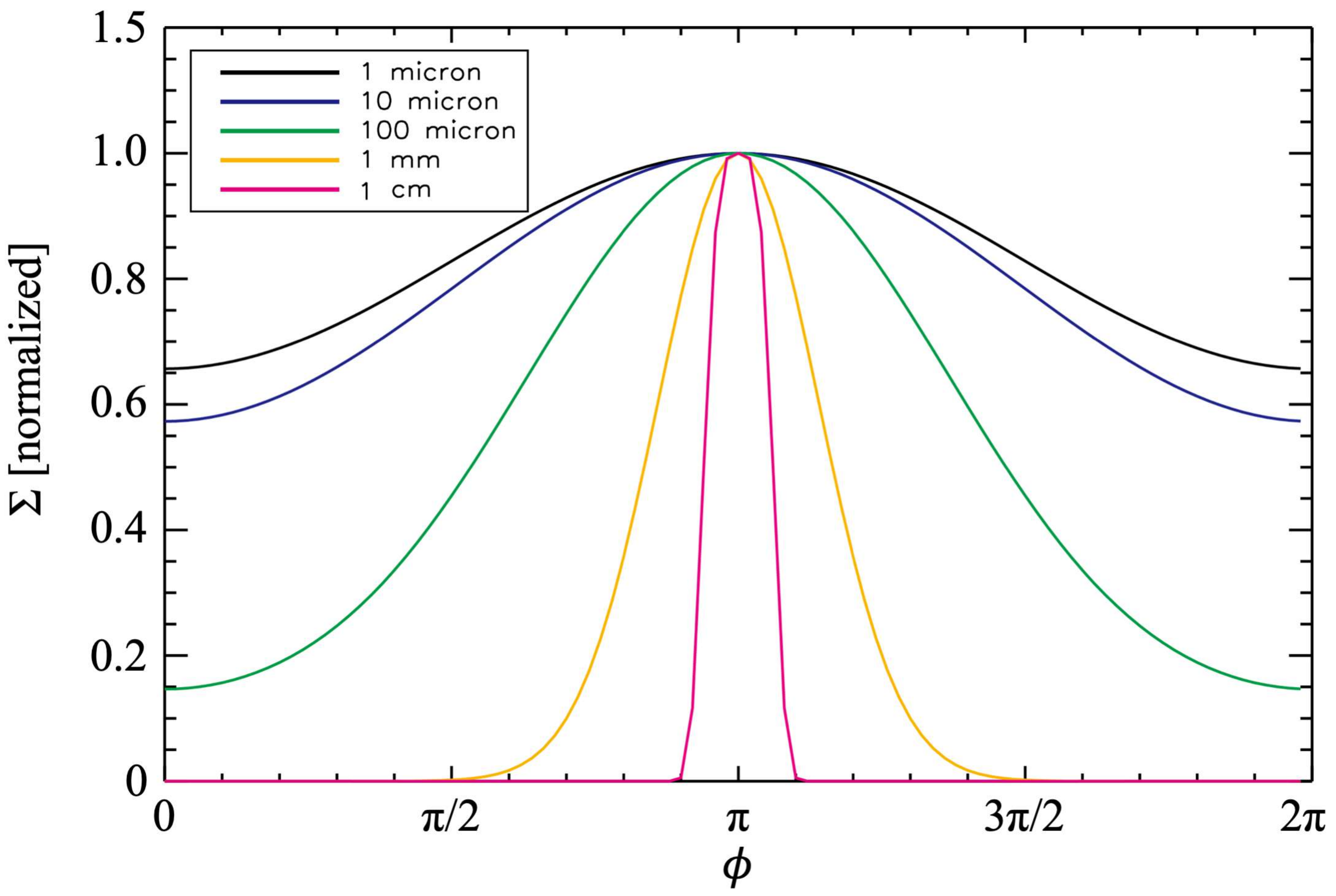}
\caption{Azimuthal distributions of dust grains for various sizes are shown, following the dust trap model \citep{bir13}.}
\label{dust_trap}
\end{figure}

By taking into account a broader grain population, we assess whether the dust trap model from \citet{bir13} is consistent with the  ALMA polarization or not.
The dust trapping model assumes that dust grains are trapped in local pressure maxima caused by a vortex \citep[e.g.,][]{bar95,kla97}. 
The distribution of dust particles shows that larger dust particles are more concentrated in the pressure maxima.
According to \citet{bir13}, the azimuthal distributions of dust grains can be expressed as
\begin{equation}
\Sigma_{\rm d}=A \Sigma_{\rm gas}\Big(-\frac{\rm St}{\alpha_{\rm t}}\Big),
\label{dust}
\end{equation}
where $A$ is a normalization constant, St is the Stokes number which is expressed as ${\rm St}=(\pi\rho a_{\rm dust})/(2\Sigma_{\rm gas})$ and $\alpha_{\rm t}$ is turbulence strength. 
In our calculations, we set a $\alpha_{\rm t}$ of $10^{-3}$.
The gas surface density is assumed to be 
\begin{equation}
\Sigma_{\rm gas}(r,\phi)=0.5 \exp\Big(\frac{-(r-r_{\rm c})^4}{2r_{\rm w}^4}\Big)\Big[1 + 0.2\sin(\phi-\frac{\pi}{2}) \Big]\, \mathrm{g\ cm^{-2}},
\label{gas}
\end{equation}
following by \citet{van16}, they derived the gas surface density of 0.5 g cm$^{-2}$ at the ring.
Then, we set the 20\% contrast of the gas surface density between the north and south parts with sinusoidal profile shown in equation (\ref{gas}). The contrast is more enhanced for larger dust grains as shown in equation (\ref{dust}).
Figure \ref{dust_trap} shows examples of azimuthal distributions of several dust grains according to equations (\ref{dust}) and (\ref{gas}).
We confirm that larger dust grains accumulate more in the dust trap.

As shown in Figure \ref{dust_trap}, we mimic the dust trapping model of \citet{bir13} from equations of (\ref{dust}) and (\ref{gas}). Here, we consider 40 populations of dust grains with different maximum grain sizes from 1 micron to $10^{5}$ micron ($=10$ cm) to perform the radiative transfer calculations. The dust size is equally spaced in log scale. 
Note that dust grains larger than 1 cm are located only in the crescent with an azimuthal width of $18^\circ$.

\begin{figure*}[htbp]
  \includegraphics[width=18.cm,bb=0 0 1534 831]{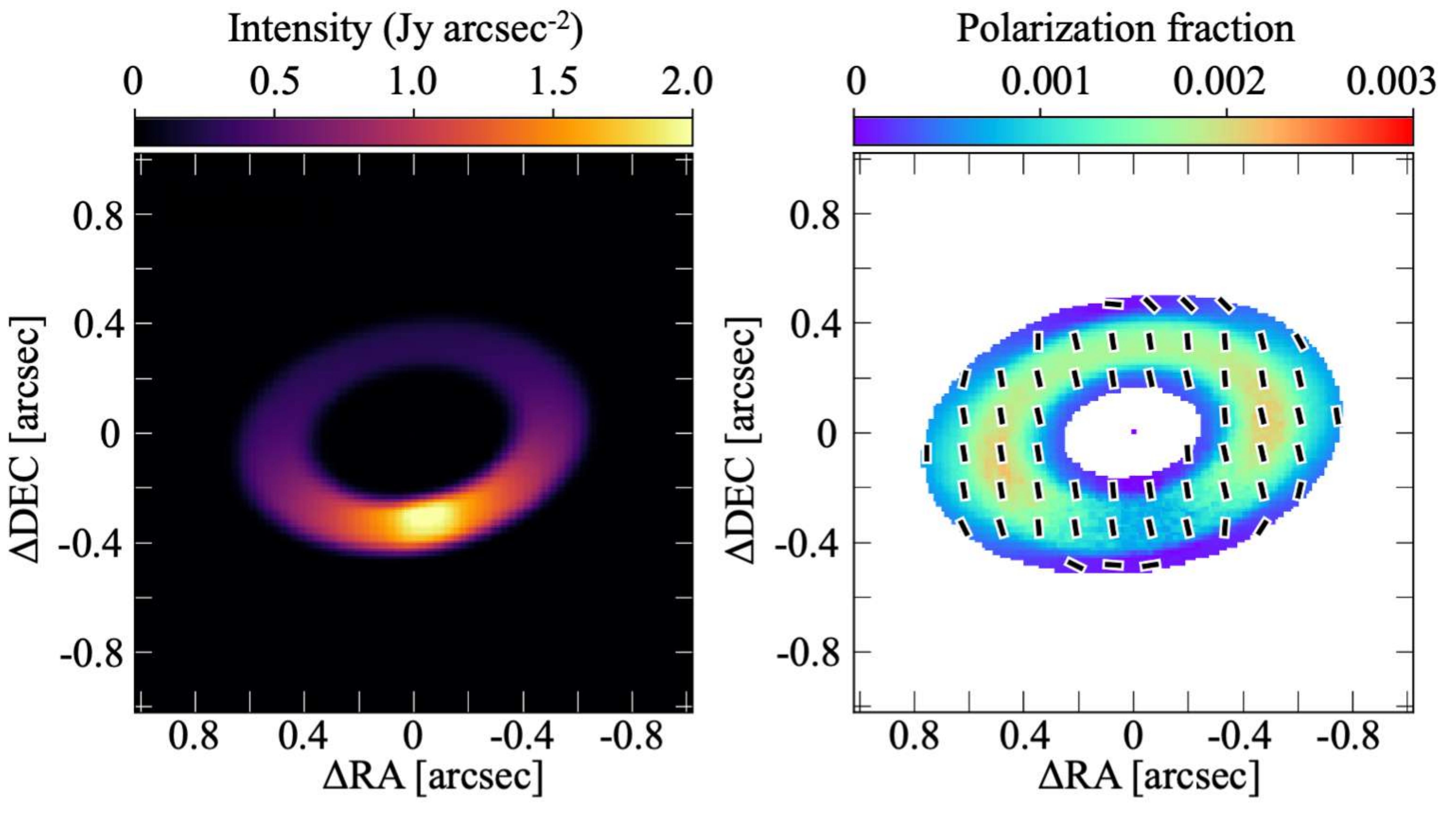}
\caption{The dust trap model of the continuum image and polarization caused by self-scattering at 860 $\mu$m wavelength (right). 40 populations of dust grains are included. }
\label{dust_trap_40}
\end{figure*}

Based on these dust distributions, we perform the radiative transfer calculations by taking into account the self-scattering effect.
Figure \ref{dust_trap_40} shows the Stokes {\it I} and polarization fraction maps of the radiative transfer calculations for the dust trapping model at 860 $\mu$m wavelength. Note that the images are not smoothed to the beam size, in order to indicate the disk structure.
The Stokes {\it I} image shows the crescent structure, similar to the ALMA observations predicted by \citet{bir13}.

However, we find that the polarization fraction only reaches $\sim0.2$\%, which is much lower than the observations. 
This is because there are many dust grains throughout the ring that contribute to the Stokes {\it I} but do not emit polarization (it should be noted that the range of the polarization fraction in Figure \ref{dust_trap_40} is much smaller than that of Figure \ref{test1}, and the color scale is different).
In detail, we find that the polarization fraction is as high as $0.21$\% on the west and east sides of the crescent, decreasing to $0.17$\% on the north part of the ring and dropping sharply to $0.06$\% at the peak of the dust trap. This latter drop in polarization fraction is because many dust grains larger than 100 $\mu$m are accumulated on the dust trap. Such larger dust grains contribute to Stokes {\it I} but do not produce the polarization.
However, the east, west, and north sides of the crescent have a population of 100 $\mu$m dust grains (see Figure \ref{dust_trap}), which contributes to both Stokes {\it I} and polarization.
Our calculations thus indicate that the dust trapping model cannot explain the ALMA polarization observations in terms of the polarization fraction at least if dust grains have already grown to larger than 100 $\mu$m in the dust trap and emission is optically thin.
We confirm this conclusion by smoothing the dust trapping model to the beam size.

In Appendix \ref{sec:Apn1}, dust trap models with 10 and 20 dust-grain populations are shown in order to investigate the effect of the number of dust populations. 
We define the 10, 20, and 40 dust-grain populations as the number of populations of different dust grain sizes.
The numbers of dust-grain populations are arbitrarily chosen.
The grain sizes are from 1 micron to $10^{5}$ micron ($=10$ cm), equally spaced in log scale as the same with the 40 dust-grain populations.
We find that these models show a polarization fraction of as low as $\sim0.2$\% and a strong decrease at the dust trap regardless of the number of dust populations.

Even though we show that the simple dust trapping model (dust grains have already grown to larger than 100 $\mu$m in the dust trap and emission is optically thin) cannot explain the ALMA polarization observation, the dust trapping might be more complex than our assumption because there will also be the grain size segregation in azimuthal direction. In this case, the  power-law distribution of grain size is suggested to be lower in the vortex \citep{sie17}.
The dust segregation localizes the 100 micron sized dust grains, resulting in lower polarization fraction \citep{poh16}, while the lower power-law distribution of grain size increases the polarization fraction.
The current polarization observation, due to low resolution, is not enough to conclude that the dust trapping model of the optically thin emission with cm dust grains is not the case.

\subsection{An Optically Thick Case}
Here we discuss an alternative scenario to explain both ALMA and VLA observations.
The lower spectral index is explained not only by grain growth but also high optical depth.
Therefore, we investigate whether an optically thick case (Model 3 in table \ref{table}) can also reproduce the observations or not.

\begin{figure*}[htbp]
  \includegraphics[width=18.cm,bb=0 0 2568 1606]{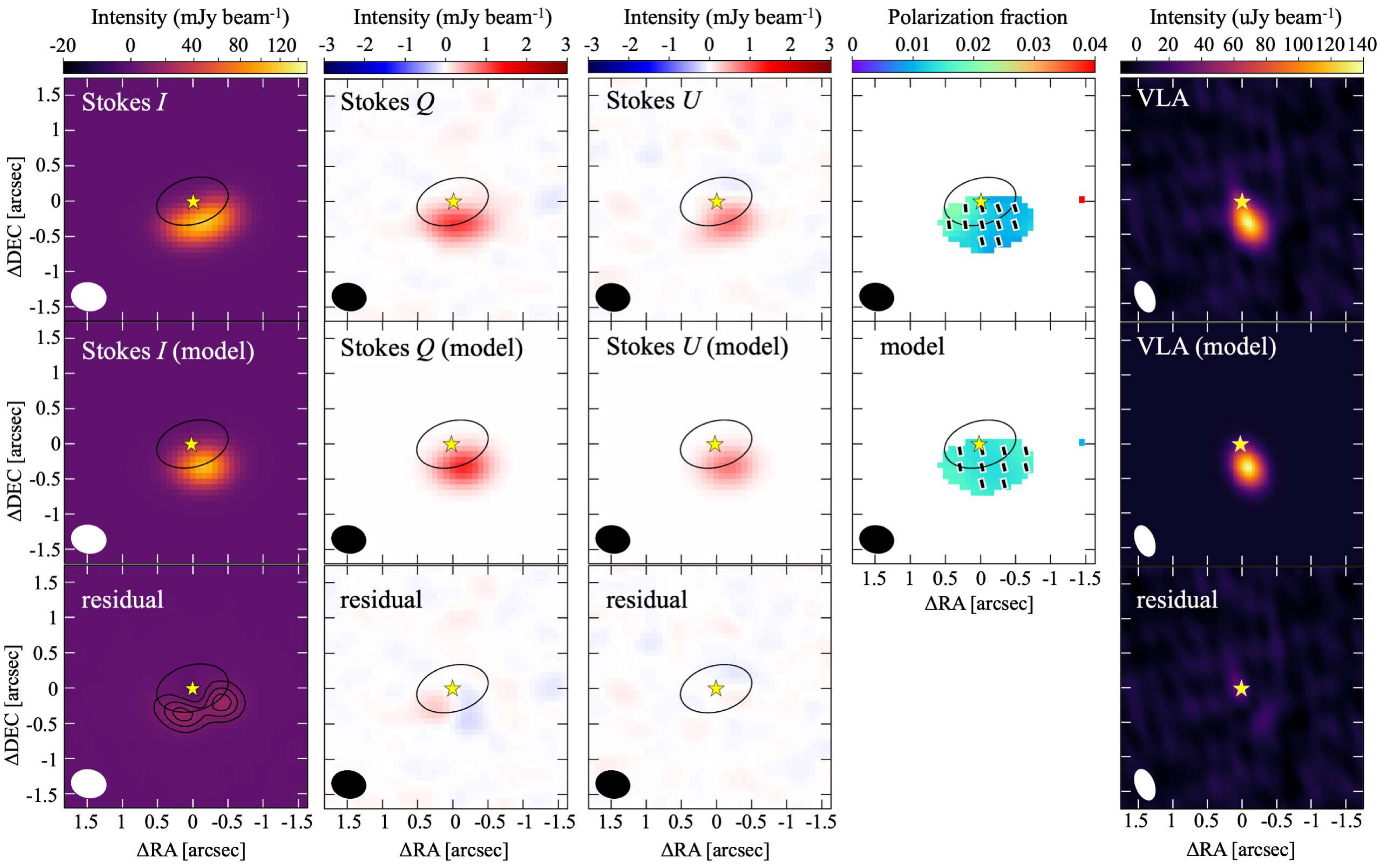}
\caption{ Same with Figure \ref{test1} but for the optically thick model assuming a maximum grain size of 140 $\mu$m, surface density of $\Sigma_{\rm d}=8.2$ g cm$^{-2}$ (corresponding to $\tau_{\rm abs}=0.09$ at $\lambda=8.8$ mm), and azimuthal width of $\phi_{\rm w}=18^\circ$. The temperature is set to be 35 K.
 }
\label{test4}
\end{figure*}

We consider the grain population with a maximum grain size of $\sim140$ $\mu$m.
We assume that the thermal emission from the 140 $\mu$m dust grains is optically thick at 860 $\mu$m, and becomes optically thin at 8.8 mm wavelength in the dust trap.
First, we reproduce the VLA continuum emission to derive the surface density of $\Sigma_0$. Then, we investigate whether this physical condition can also match the ALMA polarization data or not.

To reproduce the VLA emission, we assume an azimuthal width of $\phi_{\rm w}=18^\circ$ in equation (\ref{sigma}).
The temperature is assumed to be 35 K at 60 au in equation (\ref{temp}) because the brightness temperature was derived to be 31 K by the ALMA 440 $\mu$m dust continuum emission with higher spatial resolution \citep{van15}.
Even though the brightness temperature of the resolved continuum emission is equal to the dust temperature if the emission is optically thick,
we use a 35 K slightly higher than 31 K by taking into account the beam dilution effect. The temperature of 35 K matches the peak intensities among ALMA 440 $\mu$m, ALMA 860 $\mu$m, and VLA 8.8 mm emission (see Section \ref{two_populations}).
The previous observations derived the temperature of 60 K on the ring by assuming the axisymmetric structure \citep{bru14}. However, it may be possible for the temperature to be lowered locally, e.g. by shadowing of an inclined inner disk, as in HD 142527 \citep{cas15}.

\begin{table*}
\centering
\caption{\label{table2} Summary of the possible dust distribution models for ALMA polarization and VLA continuum}
\begin{tabular}{p{50mm}lccc}\hline\hline
Optically thin model &$\Sigma_0$$^a$ &ALMA polarization$^b$ & VLA continuum$^b$ \\\hline
1. Optically thin with  &$0.26$ g cm$^{-2}$ & yes	& no	\\
\hspace{10mm}$a_{\rm max}\sim100$ micron & & & model emission is $\sim10$ times weak \\\hline
2. Optically thin  with & $0.65$ g cm$^{-2}$ & marginally yes	& yes	\\
\hspace{5mm} $a_{\rm max}\sim100$ micron and $\sim$ cm & & Stokes {\it Q} and {\it U} drop at the continuum peak  & \\\hline\hline
Optically thick model & & ALMA polarization & VLA continuum \\\hline
3. Optically thick with & $8.2$ g cm$^{-2}$ & yes	&	yes\\
\hspace{10mm} $a_{\rm max}\sim100$ micron&  & & \\\hline
\end{tabular}
$^a$ $\Sigma_0$ is the dust surface density at the dust trap in equation (\ref{sigma}). \\
$^b$ yes/no indicates whether the model agrees with the observations or not.
\label{table2}
\end{table*}

In order to reproduce the VLA continuum emission with the 140 $\mu$m dust grains and the temperature of 35 K, the surface density needs to be $\Sigma_{0}\sim8.2$ g cm$^{-2}$ at the continuum peak.
The optical depth corresponds to $\tau_{\rm abs}=7.3$ at 860 $\mu$m and $\tau_{\rm abs}=0.090$ at 8.8 mm wavelength because the absorption opacity of the 140 $\mu$m dust grains is derived to be 0.88 g$^{-1}$ cm$^{2}$ and 0.011 g$^{-1}$ cm$^{2}$ at 860 $\mu$m and 8.8 mm wavelength, respectively, by Mie theory. Therefore, the 860 $\mu$m thermal emission is optically thick, while the 8.8 mm thermal emission is optically thin in the center of the dust trap.

Since the intensity distribution of the ALMA 440 $\mu$m emission is azimuthally wider than that of the VLA  8.8 mm emission , we also include additional dust grains with a maximum grain size of $100$ $\mu$m with an azimuthal width of $\phi_{\rm w}=41^{\circ}$ outside the optically thick region to reproduce the crescent structure observed by ALMA 440 $\mu$m continuum emission.
These additional dust grains are assumed to be optically thin at 860 $\mu$m wavelength and the emission is too faint to be observable by the VLA 8.8 mm observation.
The surface density of the additional dust grains is set to $\Sigma_{0}=0.15$ g cm$^{-2}$ in equation (\ref{sigma}) with an azimuthal width of $\phi_{\rm w}=41^{\circ}$ but is excluded in the center of the dust trap.

With these conditions ($T=35$ K and $\Sigma_0=8.2$ g cm$^{-2}$ in the center of the dust trap), we perform the radiative transfer calculation  at a wavelength of 860 $\mu$m and 8.8 mm. Then, we investigate whether these dust conditions can match the ALMA polarization and VLA continuum observations. 
Figure \ref{test4} shows the comparisons of the observations and our model for ALMA polarization and VLA continuum data.
We find that the polarization vectors are parallel to the disk minor axis and the polarization fraction is $\sim1.2-1.7$\% in the model.
The polarization fraction at the continuum peak shows 1.3\% in our model and $1.1\pm0.1$\% in the observations.
Therefore, the optically thick model matches the observations.

The model Stokes {\it Q} and {\it U} emissions have a peak emission of $\sim1$ mJy beam$^{-1}$ at the continuum peak position, matching well the observations. 
Figure \ref{test4} also shows the  model image for the VLA dust continuum emission.
We find that this optically thick model reproduces the VLA observations well.

\section{Discussion} \label{sec:discussion}

\subsection{Comparisons between our models and ALMA 440 $\mu$m dust continuum emission}\label{two_populations}

In the Section 4.1 and 4.2, we show that the ALMA polarization data and VLA continuum data would be explained by Model 2 and 3 shown in Table \ref{table2}.
Here, we discuss these two models in the context of the 440 $\mu$m dust continuum images because \citet{van13} revealed the crescent structure with ALMA 440 $\mu$m dust continuum observations with high spatial resolution.
Therefore, we calculate the radiative transfer based on these two models (Model 2 and 3) at 440 $\mu$m wavelength in order to investigate whether these models can match the ALMA 440 $\mu$m observations as well as the ALMA 860 $\mu$m and VLA 8.8 mm observations.

\begin{figure*}[htbp]
  \includegraphics[width=18.cm,bb=0 0 2932 1253]{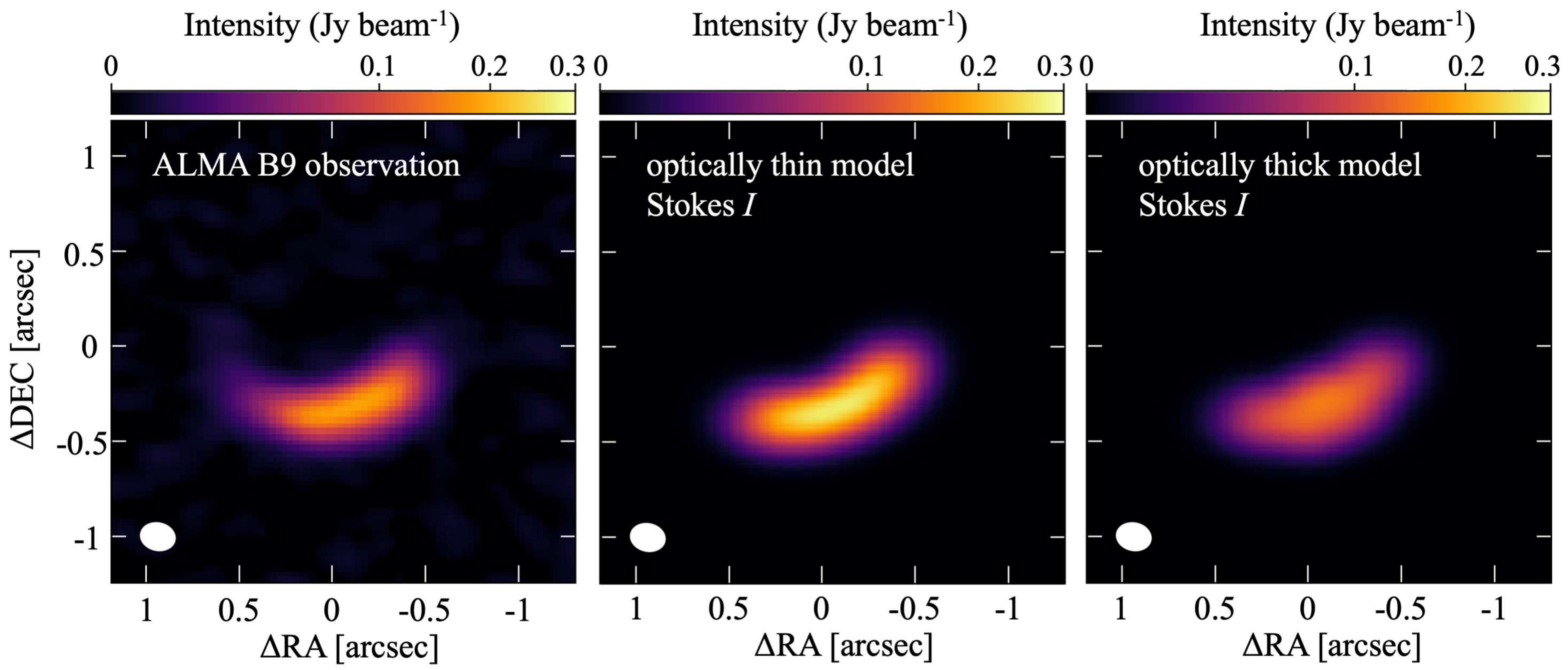}
\caption{ (Left) the ALMA 440 $\mu$m continuum image observed by \citet{van13}.  (Middle) the continuum (Stokes {\it I}) image of the optically thin model of the ALMA 440 $\mu$m observations. (Right) the continuum (Stokes {\it I}) image of the optically thick model of the ALMA 440 $\mu$m observations.}
\label{thickandthin}
\end{figure*}

\begin{figure}[htbp]
  \includegraphics[width=8.cm,bb=0 0 2243 1465]{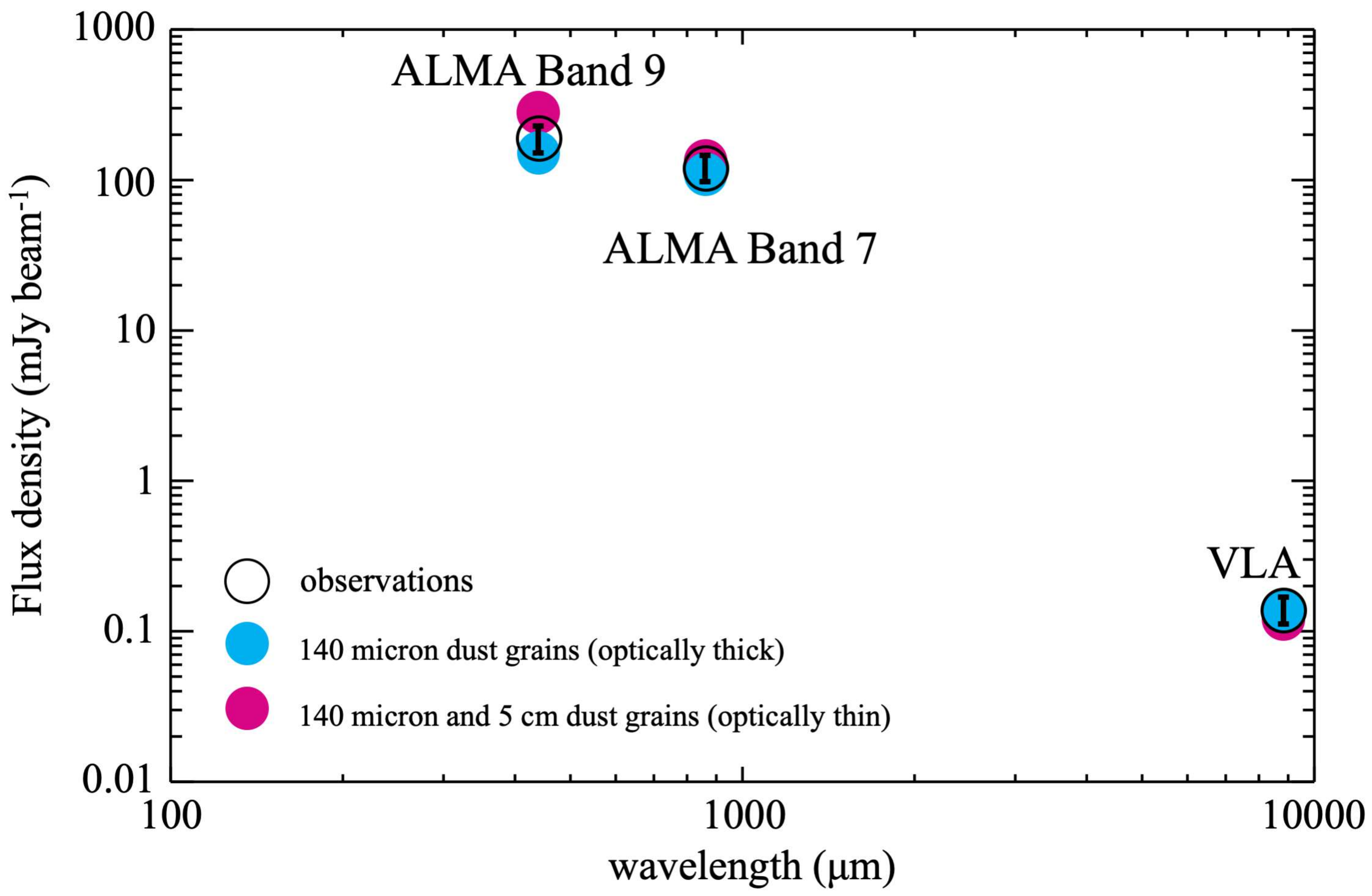}
\caption{Peak intensity against wavelength for the observations (open circles) and models (blue = 140 micron dust grains with optically thick emission, red = 140 micron and 5 cm dust grains with optically thin emission).
The flux error of the observed intensity is set to be the flux calibration uncertainty, which is 20\% for ALMA Band 9 \citep{van15}. }
\label{sed}
\end{figure}

Figure \ref{thickandthin} shows the continuum images of the ALMA 440 $\mu$m continuum observations \citep{van13}, optically thin model (Model 2 in Table \ref{table2}), and optically thick model (Model 3 in Table \ref{table2}) at 440 $\mu$m wavelength.
By comparing the observations with the model images, we find that the crescent structures of both models are similar to the observations.

We note that no emission is apparent in our models on the opposite side of the ring from the continuum emission peak even though the ALMA polarization observations detected the continuum emission with the intensity of 0.7 mJy beam$^{-1}$.
As \citet{van13} show, small dust grains would be distributed on the ring.
Such small dust grains (grain size is $<100$ $\mu$m) will contribute to the continuum emission at the northern part of the disk.
These different grain size distributions are also suggested in the HD 142527 disk by using ALMA polarization observations \citep{oha18}.

Figure \ref{sed} shows the peak intensity against wavelength for the observations and models.
The flux error of the observed intensity is set to be the flux calibration uncertainty, which is 20\% for ALMA Band 9 \citep{van15}.
The both models follow the observed peak intensities at 440 $\mu$m (ALMA Band 9), 860 $\mu$m (ALMA Band 7), and 8.8 mm (VLA) wavelengths. However, the optically thick model seems slightly better than the optically thin model in particular for the ALMA 440 $\mu$m observations.
By taking into account not only the polarization but also the continuum emission, the maximum grain size of only $\sim140$ $\mu$m in the continuum emission peak is likely.
In this case, the continuum emission at the peak is very optically thick at 860 $\mu$m wavelength ($\Sigma_{\rm d}\sim8.2$ g cm$^{-2}$ corresponds to $\tau_{\rm abs}\sim7.3$).

We point out that our proposed scenario is still consistent with the dust trapping model of \citet{bir13} in terms of the different azimuthal distributions depending on grain size. We suggest that the maximum grain size needs to be $\sim140$ $\mu$m in the dust trap for the ALMA polarization data, while \citet{bir13} assume that the dust grains are grown to centimeter size.

\subsection{Is the disk gravitationally unstable in the optically thick model?}

Here, we discuss the gravitational instability of the disk in the case that the  surface density is as high as $\Sigma_{\rm d}\sim8.2$ g cm$^{-2}$.
We use the Toomre $Q=c_{\rm s}\Omega/(\pi G\Sigma_{\rm g})$ parameter \citep{too64} to investigate whether the disk is gravitationally stable or not.
By assuming the gas to dust mass ratio of 100 and $\Sigma_{\rm d}=8.2$ g cm$^{-2}$, the Toomre $Q$ is derived to be $Q\sim0.1$, which is much lower than the critical value of $Q\sim1-2$. Therefore, the disk is gravitationally unstable to form spiral arm or fragmentation.
Conversely, the gas to dust mass ratio needs to be lower than 10 to keep it gravitationally stable ($Q\sim1$).

The requirement of low gas to dust mass ratio ($<10$) is consistent with the results of \citet{van16}, as they derive a gas surface density of 0.5 g cm$^{-2}$ at the ring. Their dust surface density   (0.04 g cm$^{-2}$) at the ring radius is not a reliable estimate as their modeling procedure considers only axisymmetric structures, i.e. this is the average dust surface density when spread over an entire ring rather than an asymmetric structure. Using a derived dust surface density of $\Sigma_{\rm d}=8.2$ g cm$^{-2}$ (in this study), the gas to dust ratio at the center of the dust trap becomes as low as 0.06. Such low gas to  dust mass ratio might cause dust feedback for destroying the vortex \citep{fu14}, although the efficiency of dust feedback in 3D is still under debate \citep{lyr18}

\begin{figure*}[htbp]
  \includegraphics[width=18.cm,bb=0 0 2999 1369]{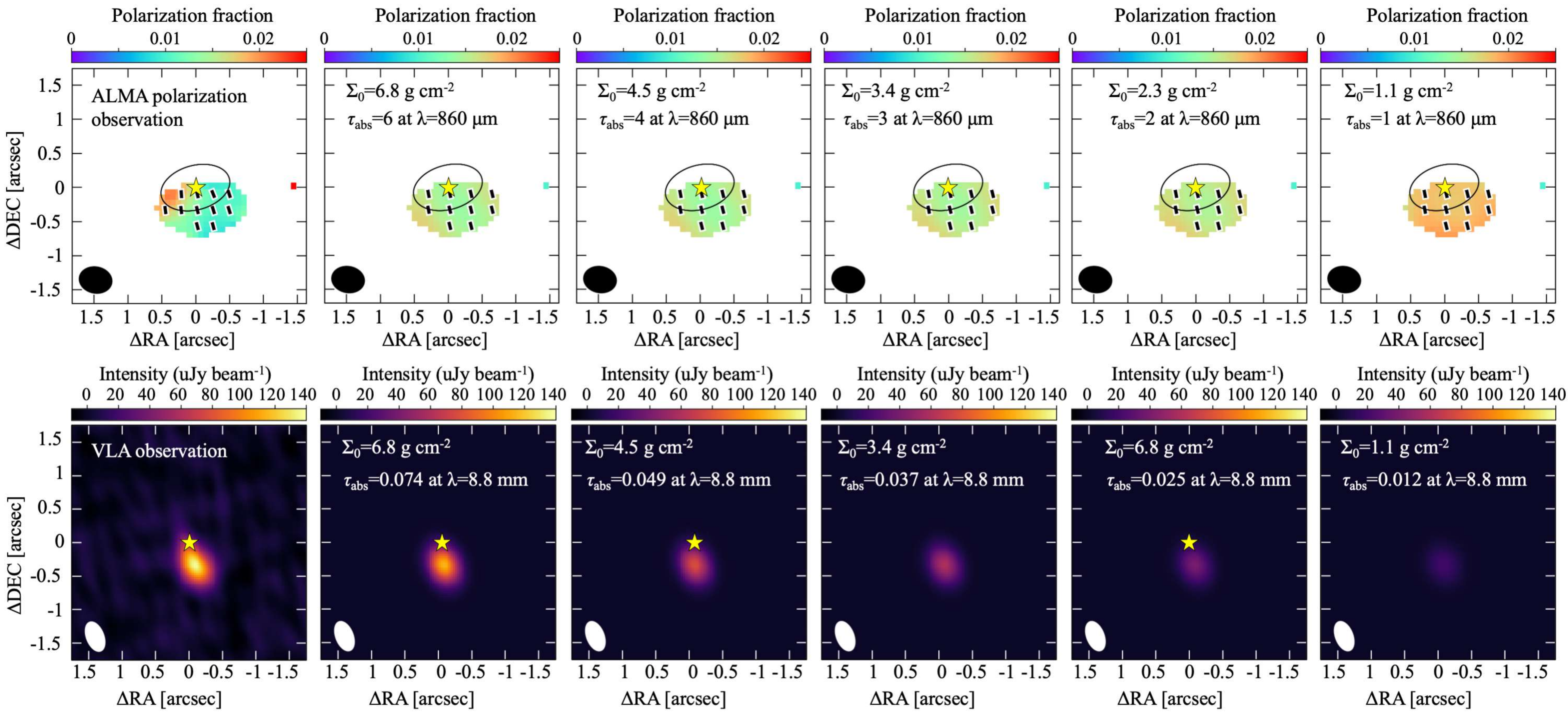}
\caption{ (Upper row) the polarization fraction overlaid with the polarization vectors of the optically thick models (Model 3 in table \ref{table2}) but for various values of the dust surface density ($\Sigma_0$), decreasing from left to right. (Lower row) the 8.8 mm dust continuum images of the same models.  The dust temperature and dust scale height are assumed to be 35 K and $f_{\rm set}=10$, respectively. The panels on the far left show the observations for the ALMA polarization and VLA continuum emission, respectively.}
\label{optical_depth}
\end{figure*}

The gas surface density measurement from the CO emission might be underestimated because of carbon depletion \citep{mio17}. The high optical depth of the dust continuum emission also affect both the molecular line emission and polarized intensity. The intensity of the molecular lines will be suppressed by the strong continuum emission, resulting in an even higher underestimate of the gas surface density. The effect of high optical depth of continuum emission on the polarization is discussed in the next section.

\subsection{Dependence of polarization and continuum emission on Optical Depth}

In the previous discussion, we show that the ALMA polarization and VLA continuum emission are explained only by 140 $\mu$m dust grains with a surface density of $\Sigma_{\rm d}=8.2$ g cm$^{-2}$ in the center of the dust trap. The optical depth corresponds to $\tau_{\rm abs}=7.3$ at 860 $\mu$m wavelength.
Here, we discuss the dependence of the ALMA polarization and VLA continuum emission on the optical depth.
Based on the  optically thick model (Model 3 in table \ref{table2}), we change the surface density ($\Sigma_0$) of the 140 $\mu$m dust grains.

Figure \ref{optical_depth} shows the effect of changing the dust surface density on the models of ALMA polarization and VLA continuum emission.
We find that the polarization fraction is saturated after the surface density is more than $\Sigma_{\rm 0}\sim3.4$ g cm$^{-2}$, corresponding to an optical depth of $\tau_{\rm abs}\sim3$ at $\lambda=860$ $\mu$m.
The saturation of the polarization fraction indicates that the polarized intensity as well as the continuum emission is observed only toward the surface layer of the disk, where the maximum grain size is 140 $\mu$m.
Therefore, the emission from the midplane of the disk cannot be observed by either continuum or polarization observations at 860 $\mu$m wavelength as the emission becomes optically thick.

In contrast, the emission of VLA 8.8 mm continuum emission remains optically thin, indicating that the VLA observations can trace the disk midplane.
Figure \ref{optical_depth} shows that the models of $1.1\ {\rm g\ cm^{-2}}\lesssim\Sigma_{\rm d} < 8.2\ {\rm g\ cm^{-2}}$ are not sufficient to reproduce the VLA observations even though these models match the ALMA polarization observations.
Therefore, it may be possible that larger dust grains are settled into the midplane, which only contribute to the VLA dust continuum emission even if the $\sim100$ $\mu$m dust grains are located at the upper layer of the disk.
Therefore, the surface density of $\Sigma_{\rm d}=8.2$ g cm$^{-2}$ can be regarded as the maximum case because the surface density higher than 8.2 g cm$^{-2}$ will exceed the VLA 8.8 mm continuum emission.

\subsection{Dependence of polarization on Dust Scale Height}\label{dust_scale}

\begin{figure}[htbp]
  \includegraphics[width=8.cm,bb=0 0 1504 1687]{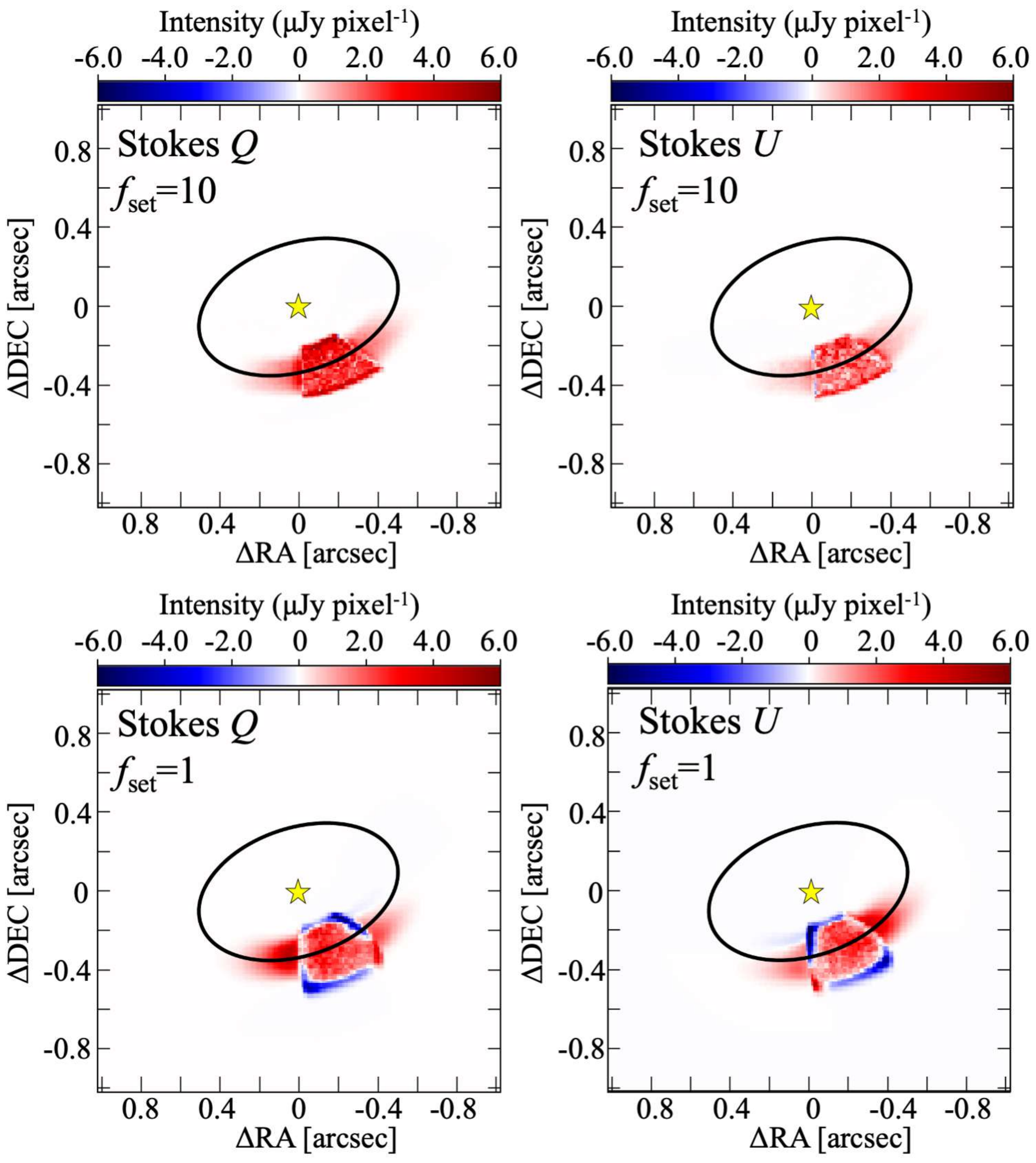}
\caption{(Upper row) Stokes {\it Q} and {\it U} images of the (optically thick) model in Figure \ref{test4} before smoothing to the beam size. The dust scale height is set to be $f_{\rm set}=10$. (Lower row) the same images as the upper panel but with the dust scale height set to $f_{\rm set}=1$. }
\label{fset1and10}
\end{figure}

\begin{figure}[htbp]
  \includegraphics[width=8.cm,bb=0 0 1496 1674]{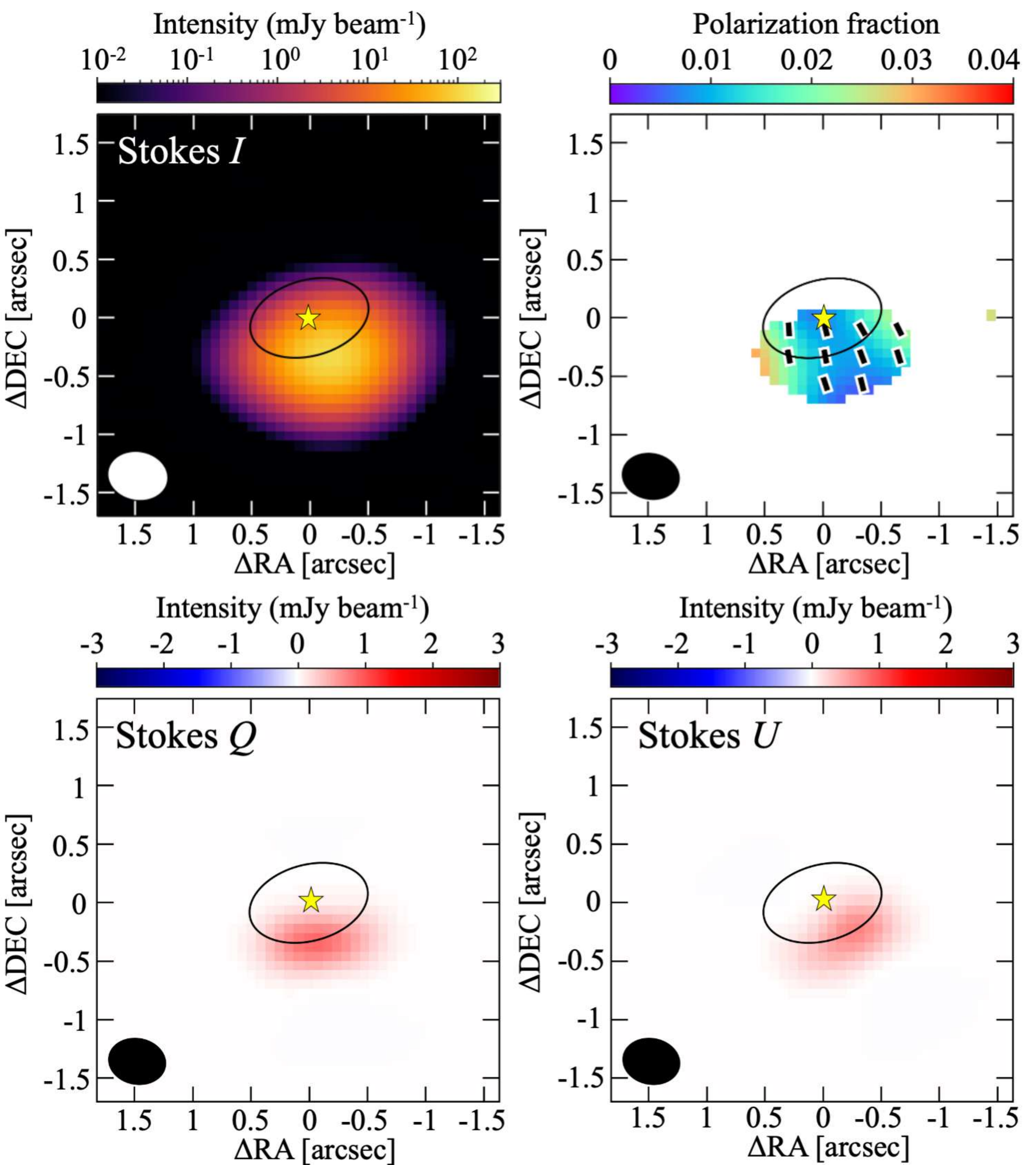}
\caption{ The optically thick model (Model 3 in Table \ref{table2}) images of the radiative transfer calculations for the total intensity (Stokes {\it I}), polarization fraction overlaid with polarization vectors, Stokes {\it Q}, and {\it U}. The dust scale height is changed to be $f_{\rm set}=1$. }
\label{test4_fset1}
\end{figure}

\begin{figure*}[htbp]
  \includegraphics[width=18.cm,bb= 0 0 2940 654]{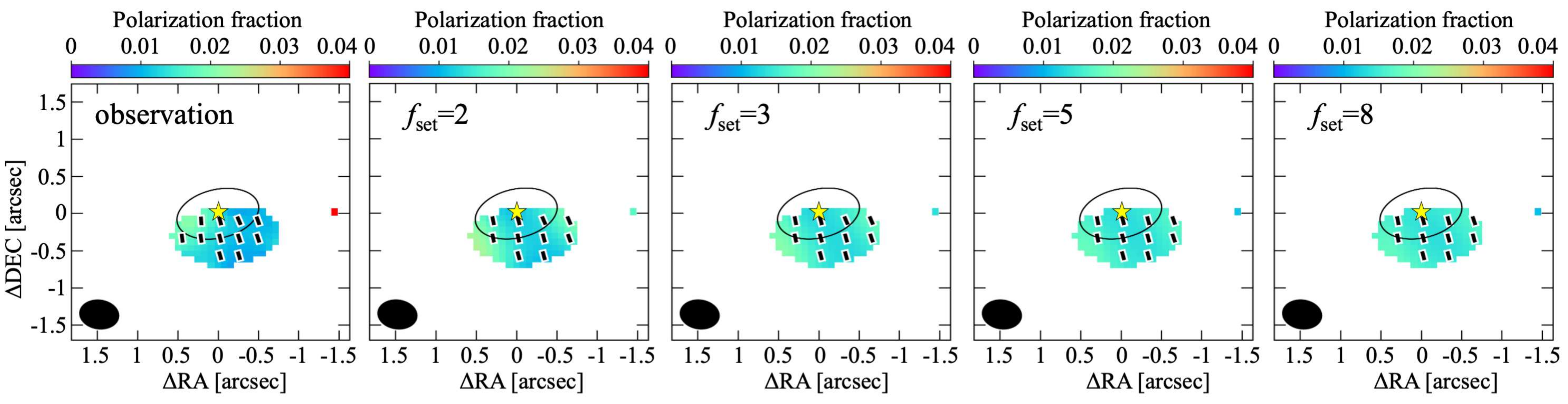}
\caption{ The polarization fraction overlaid with the polarization vectors of the models of the single grain population for various dust scale heights; the scale height decreases as $f_{\rm set}$ increases, from left to right. The models assume the maximum grain size of 140 $\mu$m, the surface density of $\Sigma_{\rm d}=8.2$ g cm$^{-2}$, and the azimuthal width of $\phi_w=18^\circ$. The temperature is set to be 35 K. The dust scale height is $f_{\rm set}=2,3,5,8$, respectively. The panel on the far left is the ALMA polarization observation. }
\label{scale_height}
\end{figure*}

In the previous models, we assume that the dust grains are well settled into the midplane.
Therefore, the dust settling parameter is set to be $f_{\rm set}=10$, which means that the dust scale height is 10 times smaller than the gas scale height.
However, \citet{oha19} pointed out that the dust scale height would change the radiation field if the surface density is fixed, which results in that the polarization patterns also change. 
Therefore, we discuss the dust scale height effect for the polarization in this disk by changing the $f_{\rm set}$ parameter.

Figure \ref{fset1and10} shows the Stokes {\it Q} and {\it U} images before smoothing to the beam size for  $f_{\rm set}=1$ and 10. 
The dust surface density is the same as the optically thick model (Model 3 in table \ref{table2}) as presented in Figure \ref{test4}.
The upper panels are the case of $f_{\rm set}=10$, and the lower panels show the case of $f_{\rm set}=1$ (the dust scale height is equal to the gas scale height).
In the case of $f_{\rm set}=10$, both of the Stokes {\it Q} and {\it U} emission only have positive values.
The Stokes {\it Q} emission is much stronger than Stokes {\it U}, indicating that the polarization vectors will be aligned with the disk minor axis.

In contrast, for the case of $f_{\rm set}=1$, we find that the Stokes {\it Q} and {\it U} emission have not only positive but also negative values in the edges of the disk.
This is the same result found by \citet{oha19} in the protoplanetary disk around HD 163296.
They pointed out that the radiation fields change if the dust scale height is changed for a fixed surface density.
Figure \ref{fset1and10} indicates that negative values of the Stokes {\it Q} and {\it U} emission are enhanced if the disk is flared.
Then, the polarization vectors are also varied.
If the flared disk is observed with large beam size such as our observation, the positive and negative emission are cancelling out, and the polarization fraction becomes lower than that of the thin disk.

Figure \ref{test4_fset1} shows the same model with Figure \ref{test4} but for $f_{\rm set}=1$, which is a beam-smoothed version of the bottom panels in Figure \ref{fset1and10}.
We find that the polarization fraction with $f_{\rm set}=1$ is $\lesssim1$\% in the dust trap, which is slightly lower than that with $f_{\rm set}=10$ and the observations.
Therefore, we prefer more dust settling, which means that vertical turbulence is less efficient.
We calculate the polarization fraction and polarization vectors by changing the dust scale height, $f_{\rm set}$.
As shown in Figure \ref{scale_height}, the polarization fraction decreases with increasing the dust scale height.
By comparing the observed polarization fraction, the dust scale height may need to be at least $f_{\rm set}\gtrsim2$ to reach the polarization fraction of $\sim1-2$\%.
However, our current polarization observations are not enough to constrain the dust scale height due to the large beam size.
Further polarization observations with higher spatial resolution and sensitivity will allow us to constrain the dust scale height.

\newpage

\section{Conclusion} \label{sec:conclusion}

We present 860 $\mu$m ALMA observations of polarized dust emission toward the Ophiuchus IRS 48 disk.
This protoplanetary disk has been also observed with ALMA 440 $\mu$m and VLA 8.8 mm dust continuum observations before and shows a lopsided ring structure.
We discuss the possible scenarios to simultaneously explain the ALMA polarization and VLA dust continuum emission.
Our main results are as follows:

\begin{itemize}
\item The polarization vectors are parallel to the disk minor axis and the polarization fraction is derived to be $1-2$\% in the southern part of the disk where polarization is detected.
These characteristics are consistent with the dust self-scattering in inclined disks.

\item We detect thermal emission from the side of the ring opposite the continuum emission peak.
The intensity is 0.7 mJy beam$^{-1}$, indicating a contrast in flux density of 170.

\item To explain both the ALMA polarization and the VLA continuum emission at 8.8 mm, we discuss two possibilities: (1) $\sim$ 5 cm dust grains are located at the center of the dust trap, whereas $\sim100$ $\mu$m dust grains are distributed outside of them, and (2) the maximum grain size is $\sim100$ $\mu$m throughout the ring, including in the dust trap. 
In the last scenario, we force the emission to be optically thick at 860 $\mu$m ($\Sigma_{\rm d}=8.2$ g cm$^{-2}$m $\tau_{\rm abs}=7.3$), while it becomes optically thin at 8.8 mm ($\tau_{\rm abs}=0.09$).
By performing radiative transfer modeling, we conclude that the optically thick model with the maximum grain size of $\sim100$ $\mu$m is more likely than the optically thin model with the two different populations of dust grain sizes.
However, it can be speculated that larger dust grains may accumulate near the midplane if the 860 $\mu$m thermal emission is optically thick.

\item In the optically thick case, we investigate the effect of the dust scale height on the polarization. We find that the polarization fraction is decreased when increasing dust scale height because the radiation field would change as shown by \citet{oha19}.
To reach a polarization fraction of $1-2$\% similar to the observations, we find that the dust scale height would be less than (1/2) $H_{\rm gas}$.

\end{itemize}

\acknowledgments
We gratefully appreciate the comments from the anonymous referee that significantly improved this article.
S.O. acknowledges supports from the Joint ALMA Observatory Visitor Program and JSPS KAKENHI Grant Numbers 18K13595 and 20K14533.
A.K. acknowledges support of JSPS KAKENHI grants 18K13590 and 19H05088.
C.L.H.H. acknowledges the support of both the NAOJ Fellowship as well as the JSPS KAKENHI grants 18K13586 and 20K14527.
P.P. acknowledges support provided by the Alexander von Humboldt Foundation in the framework of the Sofja Kovalevskaja Award endowed by the Federal Ministry of Education and Research.
This paper makes use of the following ALMA data: ADS/JAO.ALMA\#2017.1.00834.S. 
ALMA is a partnership of ESO (representing its member states), NSF (USA) and NINS (Japan), 
together with NRC (Canada), MOST and ASIAA (Taiwan), and KASI (Republic of Korea), in 
cooperation with the Republic of Chile. The Joint ALMA Observatory is operated by ESO, AUI/NRAO and NAOJ.
The National Radio Astronomy Observatory is a facility of the National Science Foundation operated under cooperative agreement by Associated Universities, Inc.

Data analysis was in part carried out on common use data analysis computer system at the Astronomy Data Center, ADC, of the National Astronomical Observatory of Japan.

\facilities{ALMA, VLA}

\software{CASA \citep[v4.5.3;][]{mcm07}, RADMC-3D \citep{dul12}
          }

%% For this sample we use BibTeX plus aasjournals.bst to generate the
%% the bibliography. The sample63.bib file was populated from ADS. To
%% get the citations to show in the compiled file do the following:
%%
%% pdflatex sample63.tex
%% bibtext sample63
%% pdflatex sample63.tex
%% pdflatex sample63.tex

%\bibliography{sample63}{}

\begin{thebibliography}{}

\bibitem[ALMA Partnership et al.(2015)]{alma15} ALMA Partnership, Brogan, C.~L., P{\'e}rez, L.~M., et al.\ 2015, \apjl, 808, L3

\bibitem[Andrews \& Williams(2005)]{and05} Andrews, S.~M., \& Williams, J.~P.\ 2005, \apj, 631, 1134

\bibitem[Andrews et al.(2016)]{and16} Andrews, S.~M., Wilner, D.~J., Zhu, Z., et al.\ 2016, \apjl, 820, L40

\bibitem[Andrews et al.(2018)]{and18} Andrews, S.~M., Huang, J., P{\'e}rez, L.~M., et al.\ 2018, \apjl, 869, L41 

\bibitem[Ansdell et al.(2018)]{ans18} Ansdell, M., Williams, J.~P., Trapman, L., et al.\ 2018, \apj, 859, 21

\bibitem[Bacciotti et al.(2018)]{bac18} Bacciotti, F., Girart, J.~M., Padovani, M., et al.\ 2018, \apjl, 865, L12 

\bibitem[Barge \& Sommeria(1995)]{bar95} Barge, P., \& Sommeria, J.\ 1995, \aap, 295, L1

\bibitem[Bertrang et al.(2018)]{ber18} Bertrang, G.~H.-M., Avenhaus, H., Casassus, S., et al.\ 2018, \mnras, 474, 5105

\bibitem[Boehler et al.(2018)]{boe18} Boehler, Y., Ricci, L., Weaver, E., et al.\ 2018, \apj, 853, 162

\bibitem[Birnstiel et al.(2010)]{bir10} Birnstiel, T., Dullemond, C.~P., \& Brauer, F.\ 2010, \aap, 513, A79

\bibitem[Birnstiel et al.(2013)]{bir13} Birnstiel, T., Dullemond, C.~P., \& Pinilla, P.\ 2013, \aap, 550, L8

\bibitem[Bohren \& Huffman(1983)]{boh83} Bohren, C.~F., \& Huffman, D.~R.\ 1983, New York: Wiley, 1983
\bibitem[Bruderer et al.(2014)]{bru14} Bruderer, S., van der Marel, N., van Dishoeck, E.~F., et al.\ 2014, \aap, 562, A26

\bibitem[Casassus et al.(2015)]{cas15} Casassus, S., Wright, C.~M., Marino, S., et al.\ 2015, \apj, 812, 126

\bibitem[Cazzoletti et al.(2018)]{caz18} Cazzoletti, P., van Dishoeck, E.~F., Pinilla, P., et al.\ 2018, \aap, 619, A161

\bibitem[Canovas et al.(2016)]{can16} Canovas, H., Caceres, C., Schreiber, M.~R., et al.\ 2016, \mnras, 458, L29

\bibitem[Casassus et al.(2013)]{cas13} Casassus, S., van der Plas, G., M, S.~P., et al.\ 2013, \nat, 493, 191


\bibitem[Cox et al.(2018)]{cox18} Cox, E.~G., Harris, R.~J., Looney, L.~W., et al.\ 2018, \apj, 855, 92 

\bibitem[Carrasco-Gonz{\'a}lez et al.(2019)]{car19} Carrasco-Gonz{\'a}lez, C., Sierra, A., Flock, M., et al.\ 2019, \apj, 883, 71

\bibitem[Cieza et al.(2017)]{cie17} Cieza, L.~A., Casassus, S., P{\'e}rez, S., et al.\ 2017, \apjl, 851, L23

\bibitem[Clarke et al.(2018)]{cla18} Clarke, C.~J., Tazzari, M., Juhasz, A., et al.\ 2018, \apjl, 866, L6

\bibitem[Dipierro et al.(2018)]{dip18} Dipierro, G., Ricci, L., P{\'e}rez, L., et al.\ 2018, \mnras, 475, 5296

\bibitem[Dong et al.(2018)]{don18} Dong, R., Liu, S.-. yuan ., Eisner, J., et al.\ 2018, \apj, 860, 124

\bibitem[Dent et al.(2019)]{den19} Dent, W.~R.~F., Pinte, C., Cortes, P.~C., et al.\ 2019, \mnras, 482, L29 

\bibitem[Dubrulle et al.(1995)]{dub95} Dubrulle, B., Morfill, G., \& Sterzik, M.\ 1995, \icarus, 114, 237 
\bibitem[Dullemond et al.(2012)]{dul12} Dullemond, C.~P., Juhasz, A., Pohl, A., et al.\ 2012, RADMC-3D: A multi-purpose radiative transfer tool, ascl:1202.015

\bibitem[Dullemond et al.(2018)]{dul18} Dullemond, C.~P., Birnstiel, T., Huang, J., et al.\ 2018, \apjl, 869, L46 

\bibitem[Draine(2006)]{dra06} Draine, B.~T.\ 2006, \apj, 636, 1114

\bibitem[Fedele et al.(2018)]{fed18} Fedele, D., Tazzari, M., Booth, R., et al.\ 2018, \aap, 610, A24

\bibitem[Flock et al.(2015)]{flo15} Flock, M., Ruge, J.~P., Dzyurkevich, N., et al.\ 2015, \aap, 574, A68

\bibitem[Fukagawa et al.(2013)]{fuk13} Fukagawa, M., Tsukagoshi, T., Momose, M., et al.\ 2013, \pasj, 65, L14

\bibitem[Fu et al.(2014)]{fu14} Fu, W., Li, H., Lubow, S., et al.\ 2014, \apjl, 795, L39

\bibitem[Gaia Collaboration et al.(2018)]{gai18} Gaia Collaboration, Brown, A.~G.~A., Vallenari, A., et al.\ 2018, \aap, 616, A1

\bibitem[Geers et al.(2007)]{gee07} Geers, V.~C., Pontoppidan, K.~M., van Dishoeck, E.~F., et al.\ 2007, \aap, 469, L35


\bibitem[Girart et al.(2018)]{gir18} Girart, J.~M., Fern{\'a}ndez-L{\'o}pez, M., Li, Z.-Y., et al.\ 2018, \apjl, 856, L27 
\bibitem[Harris et al.(2018)]{har18} Harris, R.~J., Cox, E.~G., Looney, L.~W., et al.\ 2018, \apj, 861, 91 
\bibitem[Harrison et al.(2019)]{har19} Harrison, R.~E., Looney, L.~W., Stephens, I.~W., et al.\ 2019, \apjl, 877, L2
\bibitem[Hull et al.(2018)]{hul18} Hull, C.~L.~H., Yang, H., Li, Z.-Y., et al.\ 2018, \apj, 860, 82 

\bibitem[Isella et al.(2013)]{ise13} Isella, A., P{\'e}rez, L.~M., Carpenter, J.~M., et al.\ 2013, \apj, 775, 30

\bibitem[Isella et al.(2016)]{ise16} Isella, A., Guidi, G., Testi, L., et al.\ 2016, Physical Review Letters, 117, 251101 
\bibitem[Isella et al.(2018)]{ise18} Isella, A., Huang, J., Andrews, S.~M., et al.\ 2018, \apjl, 869, L49 

\bibitem[Kraus et al.(2017)]{kra17} Kraus, S., Kreplin, A., Fukugawa, M., et al.\ 2017, \apjl, 848, L11

\bibitem[Kataoka et al.(2014)]{kat14} Kataoka, A., Okuzumi, S., Tanaka, H., \& Nomura, H.\ 2014, \aap, 568, A42 
\bibitem[Kataoka et al.(2015)]{kat15} Kataoka, A., Muto, T., Momose, M., et al.\ 2015, \apj, 809, 78 
\bibitem[Kataoka et al.(2016a)]{kat16} Kataoka, A., Muto, T., Momose, M., Tsukagoshi, T., \& Dullemond, C.~P.\ 2016a, \apj, 820, 54 
\bibitem[Kataoka et al.(2016b)]{kat16b} Kataoka, A., Tsukagoshi, T., Momose, M., et al.\ 2016b, \apj, 831, L12.
\bibitem[Kataoka et al.(2017)]{kat17} Kataoka, A., Tsukagoshi, T., Pohl, A., et al.\ 2017, \apjl, 844, L5 

\bibitem[Kenyon \& Hartmann(1987)]{ken87} Kenyon, S.~J., \& Hartmann, L.\ 1987, \apj, 323, 714

\bibitem[Kitamura et al.(2002)]{kit02} Kitamura, Y., Momose, M., Yokogawa, S., et al.\ 2002, \apj, 581, 357

\bibitem[Klahr \& Henning(1997)]{kla97} Klahr, H.~H., \& Henning, T.\ 1997, \icarus, 128, 213

\bibitem[Lyra, \& Lin(2013)]{lyr13} Lyra, W., \& Lin, M.-K.\ 2013, \apj, 775, 17

\bibitem[Lyra et al.(2018)]{lyr18} Lyra, W., Raettig, N., \& Klahr, H.\ 2018, Research Notes of the American Astronomical Society, 2, 195


\bibitem[Lee et al.(2018)]{lee18} Lee, C.-F., Li, Z.-Y., Ching, T.-C., Lai, S.-P., \& Yang, H.\ 2018, \apj, 854, 56 

\bibitem[Liu(2019)]{liu19} Liu, H.~B.\ 2019, \apjl, 877, L22

\bibitem[Lin et al.(2019)]{lin19} Lin, Z.-Y.~D., Li, Z.-Y., Yang, H., et al.\ 2019, arXiv e-prints, arXiv:1912.10012


\bibitem[Lin(2012)]{lin12} Lin, M.-K.\ 2012, \apj, 754, 21

\bibitem[Lin et al.(2019)]{2019arXiv191210012L} Lin, Z.-Y.~D., Li, Z.-Y., Yang, H., et al.\ 2019, arXiv e-prints, arXiv:1912.10012

\bibitem[Loinard et al.(2008)]{loi08} Loinard, L., Torres, R.~M., Mioduszewski, A.~J., et al.\ 2008, \apjl, 675, L29


\bibitem[Lovelace et al.(1999)]{lov99} Lovelace, R.~V.~E., Li, H., Colgate, S.~A., et al.\ 1999, \apj, 513, 805

\bibitem[Loomis et al.(2017)]{loo17} Loomis, R.~A., {\"O}berg, K.~I., Andrews, S.~M., et al.\ 2017, \apj, 840, 23

\bibitem[Long et al.(2018)]{lon18} Long, F., Pinilla, P., Herczeg, G.~J., et al.\ 2018, \apj, 869, 17


\bibitem[McMullin et al.(2007)]{mcm07} McMullin, J.~P., Waters, B., Schiebel, D., Young, W., \& Golap, K.\ 2007, Astronomical Data Analysis Software and Systems XVI, 376, 127 

\bibitem[Mathis et al.(1977)]{mat77} Mathis, J.~S., Rumpl, W., \& Nordsieck, K.~H.\ 1977, \apj, 217, 425 
\bibitem[Miyake \& Nakagawa(1993)]{miy93} Miyake, K., \& Nakagawa, Y.\ 1993, \icarus, 106, 20 

\bibitem[Miotello et al.(2017)]{mio17} Miotello, A., van Dishoeck, E.~F., Williams, J.~P., et al.\ 2017, \aap, 599, A113



\bibitem[Mori et al.(2019)]{mor19} Mori, T., Kataoka, A., Ohashi, S., et al.\ 2019, arXiv e-prints, arXiv:1907.10229
\bibitem[Nagai et al.(2016)]{nag16} Nagai, H., Nakanishi, K., Paladino, R., et al.\ 2016, \apj, 824, 132 
\bibitem[Ohashi et al.(2018)]{oha18} Ohashi, S., Kataoka, A., Nagai, H., et al.\ 2018, \apj, 864, 81
\bibitem[Ohashi \& Kataoka(2019)]{oha19} Ohashi, S., \& Kataoka, A.\ 2019, \apj, 886, 103

\bibitem[Ono et al.(2016)]{ono16} Ono, T., Muto, T., Takeuchi, T., et al.\ 2016, \apj, 823, 84

\bibitem[Okuzumi, \& Tazaki(2019)]{oku19} Okuzumi, S., \& Tazaki, R.\ 2019, \apj, 878, 132

\bibitem[P{\'e}rez et al.(2014)]{per14} P{\'e}rez, L.~M., Isella, A., Carpenter, J.~M., et al.\ 2014, \apjl, 783, L13

\bibitem[Pollack et al.(1994)]{pol94} Pollack, J.~B., Hollenbach, D., Beckwith, S., et al.\ 1994, \apj, 421, 615 
\bibitem[Pohl et al.(2016)]{poh16} Pohl, A., Kataoka, A., Pinilla, P., et al.\ 2016, \aap, 593, A12 

\bibitem[P{\'e}rez et al.(2014)]{per14} P{\'e}rez, L.~M., Isella, A., Carpenter, J.~M., et al.\ 2014, \apjl, 783, L13

\bibitem[Pinte et al.(2016)]{pin16} Pinte, C., Dent, W.~R.~F., M{\'e}nard, F., et al.\ 2016, \apj, 816, 25


\bibitem[Sheehan, \& Eisner(2018)]{she18} Sheehan, P.~D., \& Eisner, J.~A.\ 2018, \apj, 857, 18

\bibitem[Ricci et al.(2010)]{ric10} Ricci, L., Testi, L., Natta, A., et al.\ 2010, \aap, 512, A15
\bibitem[Ricci et al.(2012)]{ric12} Ricci, L., Trotta, F., Testi, L., et al.\ 2012, \aap, 540, A6

\bibitem[Sadavoy et al.(2018)]{sad18} Sadavoy, S.~I., Myers, P.~C., Stephens, I.~W., et al.\ 2018, \apj, 859, 165 
\bibitem[Sadavoy et al.(2019)]{sad19} Sadavoy, S.~I., Stephens, I.~W., Myers, P.~C., et al.\ 2019, \apjs, 245, 2

\bibitem[Schr{\"a}pler \& Henning(2004)]{sch04} Schr{\"a}pler, R., \& Henning, T.\ 2004, \apj, 614, 960


\bibitem[Stephens et al.(2014)]{ste14} Stephens, I.~W., Looney, L.~W., Kwon, W., et al.\ 2014, \nat, 514, 597
\bibitem[Stephens et al.(2017)]{ste17} Stephens, I.~W., Yang, H., Li, Z.-Y., et al.\ 2017, \apj, 851, 55 

\bibitem[Sierra et al.(2017)]{sie17} Sierra, A., Lizano, S., \& Barge, P.\ 2017, \apj, 850, 115


\bibitem[Tazaki et al.(2019)]{taz19} Tazaki, R., Tanaka, H., Kataoka, A., et al.\ 2019, \apj, 885, 52

\bibitem[Testi et al.(2014)]{tes14} Testi, L., Birnstiel, T., Ricci, L., et al.\ 2014, Protostars and Planets VI, 339

\bibitem[Tsukagoshi et al.(2016)]{tsu16} Tsukagoshi, T., Nomura, H., Muto, T., et al.\ 2016, \apjl, 829, L35

\bibitem[Toomre(1964)]{too64} Toomre, A.\ 1964, \apj, 139, 1217

\bibitem[Ueda et al.(2020)]{ued20} Ueda, T., Kataoka, A., \& Tsukagoshi, T.\ 2020, arXiv e-prints, arXiv:2003.09353

\bibitem[van der Marel et al.(2013)]{van13} van der Marel, N., van Dishoeck, E.~F., Bruderer, S., et al.\ 2013, Science, 340, 1199
\bibitem[van der Marel et al.(2015)]{van15} van der Marel, N., Pinilla, P., Tobin, J., et al.\ 2015, \apjl, 810, L7
\bibitem[van der Marel et al.(2016)]{van16} van der Marel, N., van Dishoeck, E.~F., Bruderer, S., et al.\ 2016, \aap, 585, A58

\bibitem[van der Marel et al.(2019)]{van19} van der Marel, N., Dong, R., di Francesco, J., et al.\ 2019, \apj, 872, 112

\bibitem[van der Plas et al.(2017)]{van17} van der Plas, G., Wright, C.~M., M{\'e}nard, F., et al.\ 2017, \aap, 597, A32

\bibitem[van Terwisga et al.(2018)]{van18} van Terwisga, S.~E., van Dishoeck, E.~F., Ansdell, M., et al.\ 2018, \aap, 616, A88


\bibitem[Warren(1984)]{war84} Warren, S.~G.\ 1984, \ao, 23, 1206 
\bibitem[Weingartner \& Draine(2001)]{wei01} Weingartner, J.~C., \& Draine, B.~T.\ 2001, \apj, 548, 296 
\bibitem[Yang et al.(2016)]{yang16} Yang, H., Li, Z.-Y., Looney, L., \& Stephens, I.\ 2016, \mnras, 456, 2794 
\bibitem[Yang et al.(2017)]{yang17} Yang, H., Li, Z.-Y., Looney, L.~W., Girart, J.~M., \& Stephens, I.~W.\ 2017, \mnras, 472, 373 

\bibitem[Youdin \& Lithwick(2007)]{you07} Youdin, A.~N., \& Lithwick, Y.\ 2007, \icarus, 192, 588 

\bibitem[Zhu, \& Stone(2014)]{zhu14} Zhu, Z., \& Stone, J.~M.\ 2014, \apj, 795, 53

\bibitem[Zhu et al.(2019)]{zhu19} Zhu, Z., Zhang, S., Jiang, Y.-F., et al.\ 2019, \apjl, 877, L18

\end{thebibliography}

%\bibliographystyle{aasjournal}

\appendix\section{Dust trap model for ALMA polarization with 10 and 20 dust-grain populations }\label{sec:Apn1}

Here, we investigate the dust trap model with 10 and 20 dust-grain populations because the number of duet grains in size may affect the polarization fraction.
The model with 40 dust populations is shown in Section \ref{sec:dust_trap}.

\begin{figure}[htbp]
  \includegraphics[width=18.cm,bb=0 0 1561 834]{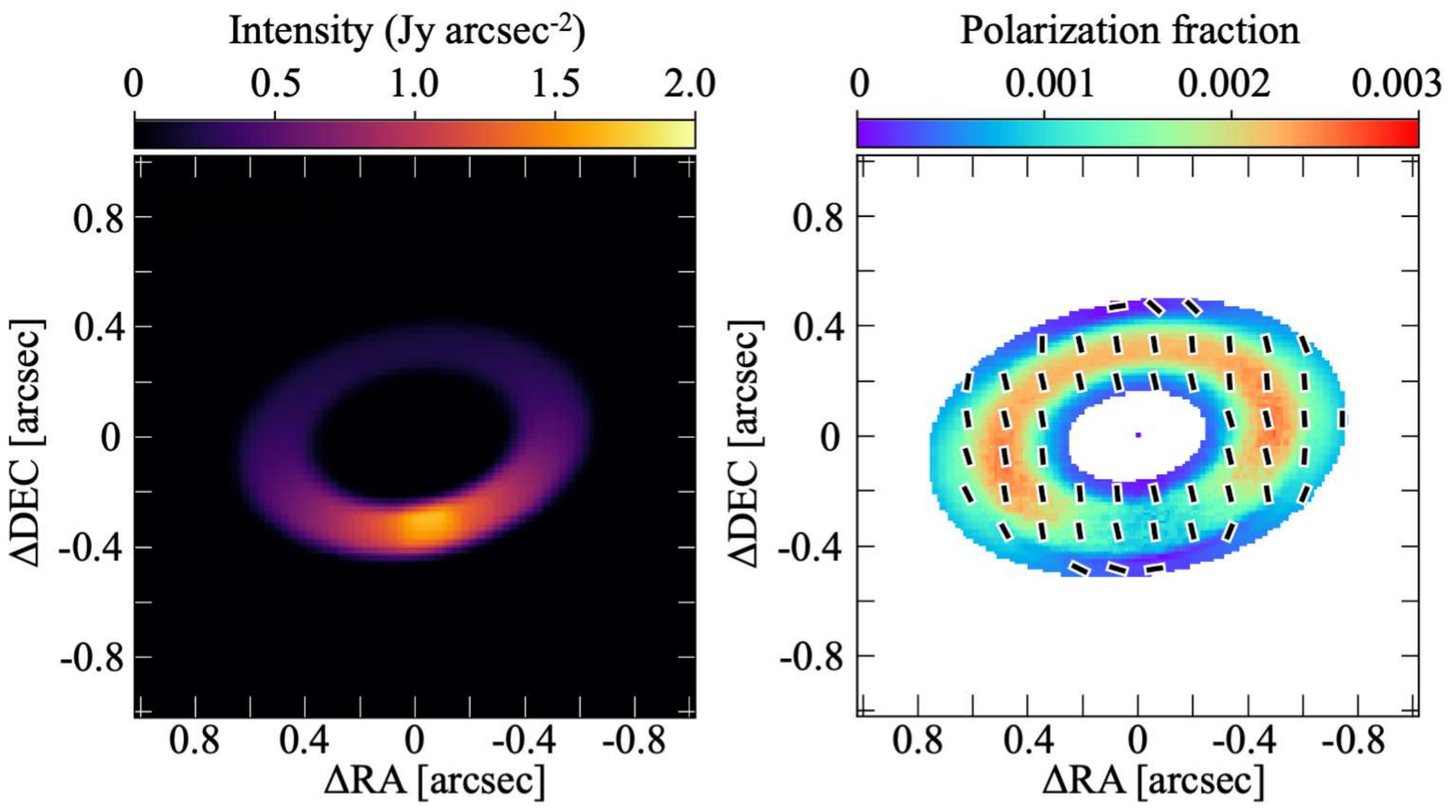}
\caption{The dust trap model of the continuum image and polarization fraction caused by self-scattering at 860 $\mu$m wavelength. The 10 populations of dust grains are included. }
\label{dust_trap_10}
\end{figure}

\begin{figure}[htbp]
  \includegraphics[width=18.cm,bb=0 0 1561 829]{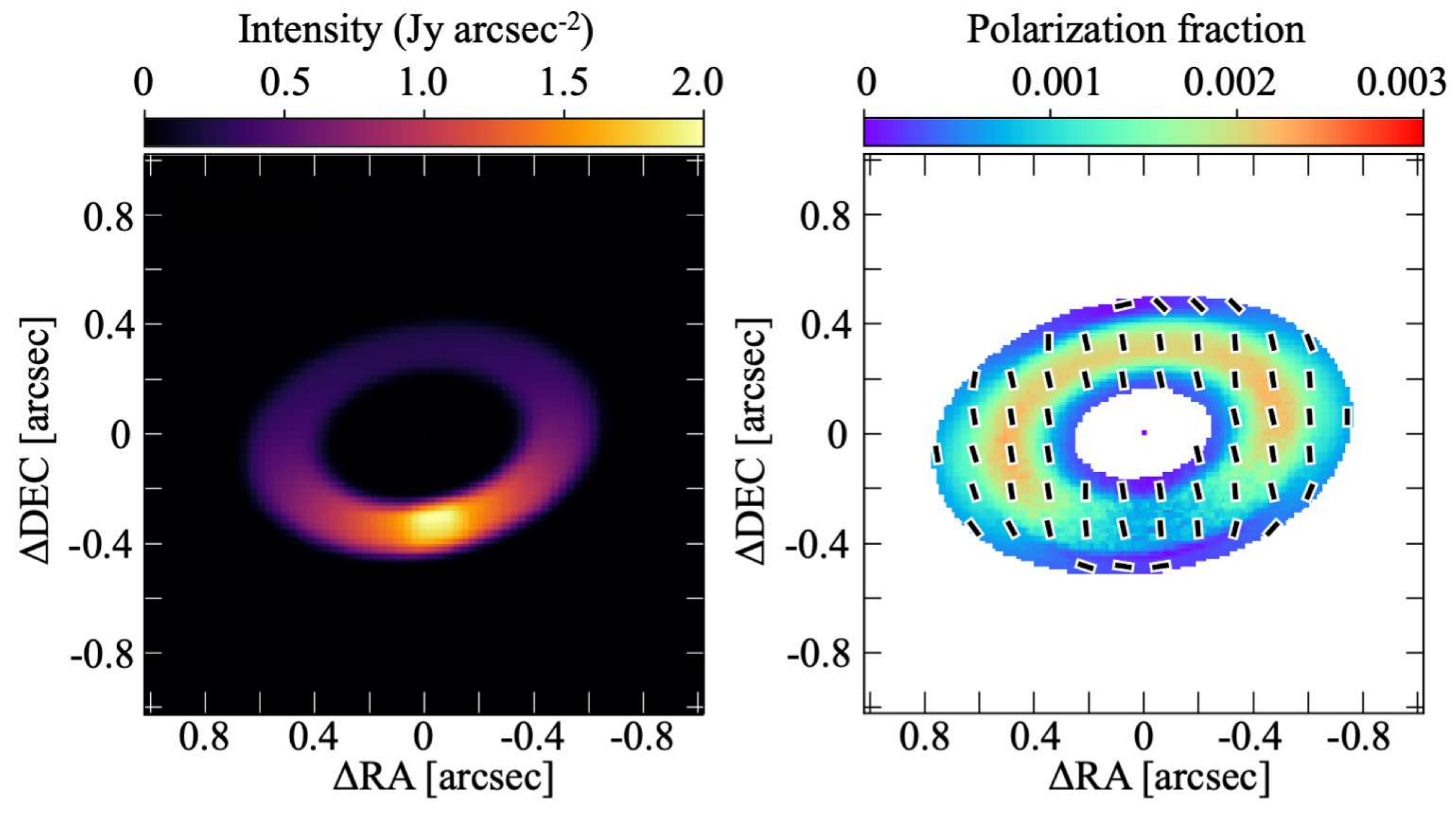}
\caption{The dust trap model same with Figure \ref{dust_trap_10} but the 20 populations of dust grains are included.}
\label{dust_trap_20}
\end{figure}

Figure \ref{dust_trap_10} and \ref{dust_trap_20} shows the cases where 10 and 20 dust-grains populations are included, respectively. We find that these models show a polarization fraction of as low as $\sim0.2$\% and drops at the dust trap regardless of the number of dust populations, which indicates that the surface density the of $\sim100$ micron sized dust grains are not enough to produce the observed polarization fraction.
Therefore, we conclude that the dust trapping model cannot explain the ALMA polarization observations in the optically thin case.

%% This command is needed to show the entire author+affiliation list when
%% the collaboration and author truncation commands are used.  It has to
%% go at the end of the manuscript.
%\allauthors

%% Include this line if you are using the \added, \replaced, \deleted
%% commands to see a summary list of all changes at the end of the article.
%\listofchanges

\end{document}